\documentclass[final,3p,authoryear]{elsarticle} 

\usepackage{amsfonts}
\usepackage{amsmath}
\usepackage{amssymb}
\usepackage{amsthm}
\usepackage{bm}
\usepackage{graphicx,psfrag,epsf}
\usepackage{enumerate}
\usepackage{enumitem}
\usepackage{natbib}
\usepackage{url} 
\usepackage{subfigure}
\usepackage{mathtools}
\usepackage{textcomp} 
\usepackage{color}

\usepackage{tabularx}
\usepackage{rotating}

\usepackage{amsthm}

\usepackage{amsthm}

\theoremstyle{remark}
\newtheorem{remark}{Remark}

\theoremstyle{definition}

\theoremstyle{remark}

\newcolumntype{L}[1]{>{\hsize=#1\hsize\raggedright\arraybackslash}X}%

\usepackage{todonotes}

\usepackage{setspace}
\newcommand{\E}{\text{E}}
\newcommand{\Var}{\text{Var}}

\newcommand{\argmax}{\operatornamewithlimits{argmax}}

\theoremstyle{definition}

\usepackage{ccaption}

\journal{ADAC - 
Special Issue on ``Advances in Clustering, Classification and Related Methods''}

\biboptions{authoryear}

\begin{document}

\begin{frontmatter}

\title{Estimation and Inference for Multiplicative Error Models via Analytical Derivatives}
\author[Luca]{Luca Scaffidi Domianello\corref{cor}} 
\ead{luca.scaffidi@unict.it}
\cortext[cor]{Luca Scaffidi Domianello – Dipartimento di Economia e Impresa, Università di Catania, Corso Italia 55, 95129, Catania, Italy.}
\address[Luca]{Dipartimento di Economia e Impresa, Università di Catania, Italy}

\begin{abstract}
Maximum Likelihood Estimates (MLEs) of the parameters of a Multiplicative Error Model are typically computed through a standard gradient method, an iterative procedure usually based on numerical approximations to derivatives of the log-likelihood function, owing to the complexity of the analytical derivatives, which involve recursive terms. Building on previous work developed mainly in the GARCH framework, we derive analytical expressions for the score vector and the Hessian matrix of the Multiplicative Error Model log-likelihood by adopting the Gamma distribution for estimation purposes. Then, we illustrate how the iterative algorithm based on the analytical derivatives should be computationally implemented. After maximizing the log-likelihood, we also provide explicit formulas for several estimators of the asymptotic covariance matrix of the MLE. Additionally, we provide analytic formulas for robust covariance estimators that remain valid under distributional misspecification, which are also referred to as the asymptotic variance of the Quasi-Maximum Likelihood Estimator. We also propose a covariance estimator that is consistent under general forms of heteroskedasticity of the error term, while the usually adopted estimators rely on a specific variance assumption. The finite-sample performance of the covariance estimators is assessed through a Monte Carlo investigation. An application on real data completes the analysis.
\end{abstract}

\end{frontmatter}


\newpage

\section{Introduction}
\label{sec:Introduction}

Financial volatility modeling within the GARCH framework is based on daily squared returns, a noisy, albeit unbiased, proxy of the latent conditional variance. Nevertheless, the availability of ultra-high frequency data (UHFD) in the last two decades, allows for more a precise volatility proxy within the wide class of realized volatility \citep[RV,][]{Andersen:Bollerslev:Diebold:Labys:2003}: from the plain vanilla version of RV, as the square root of the sum of squared high-frequency returns (sampled at, e.g., five minutes), more refined versions taking into account autocorrelation of intradaily returns or microstructure noise, such as the realized kernel volatility of \citet{BarndorffNielsen:Hansen:Lunde:Shephard:2008}, are available as more robust estimators of volatility. 
Additionally, following \citet{Andersen:Bollerslev:1998}, UHFD-based volatility measures are used as the suitable target to evaluate the forecasting performance of volatility models, due to the latent nature of time-varying variance, in place of squared returns.

UHFD-based volatility measures exhibit some empirical regularities, which are commonly referred to as \textit{stylized facts}. One of the main features is the persistence of the series, said differently, the RV series is highly autocorrelated. Hence, researchers in the last two decades have specified dynamics for RV measures directly, thus treating volatility as an observable quantity rather than a latent variable, like in the GARCH framework. Furthermore, being volatility highly persistent, a more accurate measure of volatility is valuable for forecasting purposes \citep{Hansen:2011}. As a consequence, new models such as the Multiplicative Error Model \citep[MEM;][]{Engle:2002,Engle:Gallo:2006} or the Heterogeneous AutoRegressive model \citep[HAR;][]{Corsi:2009}  have proved useful to mimic its dynamics, and several extensions have been proposed in the literature.
The estimation of MEM is usually performed through the Maximum Likelihood method by adopting the conditional Gamma distribution for the RV measure. Maximum Likelihood Estimates (MLEs) are usually computed by iterating a gradient-based algorithm until convergence, since closed-form solutions are not available. Moreover, the derivatives of the log-likelihood, given their recursive structure, are usually computed through numerical differentiation. Nevertheless, it is well known that numerical derivatives are unstable, especially when they are used to compute the Hessian matrix. On the contrary, the use
of analytical derivatives in the estimation procedure should improve the accuracy of the resulting estimates considerably \citep{Mccullough:1999}.
For this purpose, \cite{Fiorentini:1996} provide analytical derivatives of the GARCH log-likelihood up to the second order. Furthermore, relying on these analytical expressions, they compare the behaviour of several asymptotic covariance matrix estimators.
To the best of our knowledge, such analytical derivatives, as well as closed-form expressions for the different estimators of the covariance matrix, are not available for the MEM framework.
First, in our work, we provide analytical expressions for the score and the Hessian matrix of the MEM log-likelihood by adopting the Gamma distribution for the errors for estimation purposes. In principle, any distribution defined on a positive support can be adopted;\footnote{Lognormal, Weibull, Exponential, Inverse Gaussian, just to name a few.} nevertheless, the Gamma distribution, given its Quasi Maximum Likelihood properties for the conditional mean equation parameters, is widely used in empirical works. \\
Second, based on explicit formulas of the score and the Hessian matrix, we compute several estimators of the asymptotic covariance matrix, and we compare their finite sample behaviour, under different conditions (e.g., correct specification and misspecification in distribution), through a simulation study.\\
Third, we illustrate how to implement an iterative algorithm, based on the analytical expressions we derived, which is readily programmable, to compute the MLEs of the model parameters. On the contrary, most software packages available rely on numerical approximations to derivatives, which are sometimes unstable. \\
Our simulation results show that the usually adopted estimator of the asymptotic covariance matrix, based on a specific conditional variance assumption, is not consistent when, as expected, the latter does not hold, while the estimator we propose is consistent general forms of heteroskedasticity of the error term.  Moreover, the covariance assumption has implications for estimating the asymptotic covariance matrix of the estimated parameters, while it is not required to get consistent estimates of the conditional mean equation parameters; this is something that has not been emphasized in the MEM literature. As a consequence, as underlined in the real case study, we suggest reporting both estimated standard errors for comparison.\\
The paper is structured as follows: Section \ref{sec:MEM}
provides a brief description of the MEM and its extensions, while Section \ref{sec:est} illustrates the MLEs of the model parameters (Section \ref{sec:mle}), the iterative algorithm adopted to compute the MLE (Section \ref{sec:scoring}), and the Quasi Maximum Likelihood interpretation of the MLE along with the corresponding estimators of the asymptotic covariance matrix (Section \ref{sec:qmle}). Section \ref{sim} describes the simulation study, under both correct specification and misspecification in distribution, while  In Section \ref{sec:empr},  we estimate the two most used specifications of the MEM for the 30 constituent stocks of the Dow Jones Industrial Average Index (DJIA) by using the iterative algorithm illustrated throughout the text. Finally, Section \ref{sec:concl} provides some concluding remarks.

\section{The Multiplicative Error Model (MEM)}
\label{sec:MEM}

Let $\left\{x_{t}\right\}$ be a non--negative time series process whose dynamics is described by the following Multiplicative Error Model, hereafter denoted by MEM($P$,$Q$) 
\begin{align}
	&x_{t}= \mu_t \varepsilon_{t}, \ \qquad
	\varepsilon _{t}\stackrel{i.i.d.}\sim \mathcal{D}^+\left(1,\phi^2  \right), \label{errorMEM} \\
	&\E(x_t|\mathcal{I}_{t-1})	=\mu_{t}= \omega + \sum_{p=1}^P\alpha_p x_{t-p}+ \sum_{q=1}^Q\beta_q \mu_{t-q},
	\label{eq:mean-MEM}	\\ 
	&\Var(x_t|\mathcal{I}_{t-1})=\phi^2\mu_t^2, \label{condvar}
\end{align} 
where $\left\lbrace\varepsilon_{t}\right\rbrace$ is a sequence of independent and identically distributed (i.i.d.) random variables, defined on a positive support, with a unit mean and finite variance, $\phi^2>0$,\footnote{In the Generalized Linear Model (GLM) literature, $\phi^2$ is referred to as the dispersion parameter.} while $P$ and $Q$ denote the number of lags of the dependent variable and its conditional mean, respectively. Usually, $Q$ is set to 1, while $P$ is equal to 1 or 2.
Given the unit expected value of $\varepsilon_t$, a usual assumption in the MEM literature \citep[see, e.g.,][]{Brownlees:2012}, the expectation of $x_t$ conditionally on the information set at the previous period, $\mathcal{I}_{t-1}=\left\lbrace x_{t-1},\dots,x_{1}\right\rbrace$, is equal to $\mu_{t}$, i.e. $\E\left( x_{t}\rvert\mathcal{I}_{t-1}\right)=\mu_{t}$. Conditionally on the information set available at date $t-1$, $\mu_t$, as specified in \eqref{eq:mean-MEM}, evolves deterministically, where $\alpha_p$ is the impact of the $p$th lag of $x_t$ on $\mu_t$, while $\beta_q$ measures the inertial contribution of $\mu_{t-q}$ to the conditional mean at time $t$.

Starting from the specification of $\mu_t$ in \eqref{eq:mean-MEM}, we can derive an ARMA($M$,$Q$) representation for $x_t$, with $M=\max\left\lbrace P,Q\right\rbrace$. By rearranging \eqref{eq:mean-MEM}, we get
\begin{equation}
	x_t=\omega +  \sum_{p=1}^{\max\left\lbrace P,Q\right\rbrace}(\alpha_p+\beta_p) x_{t-p} -  \sum_{q=1}^Q\beta_q  u_{t-q} + u_t,
\end{equation}
where $u_t=\mu_t(\varepsilon_t-1)$ is a martingale difference sequence, then $\E(u_t|\mathcal{I}_{t-1})=0$. Nevertheless, $u_t$ is clearly conditionally heteroskedastic, given the multiplicative structure of the disturbances, $\varepsilon_t\text{'s}$.\footnote{Note that $\Var(u_t|\mathcal{I}_{t-1})=\phi^2\mu_t^2$.} 

According to \eqref{errorMEM}-\eqref{condvar}, $x_t|\mathcal{I}_{t-1}\sim \mathcal{D}^+\left(\mu_t, \phi^2 \mu^2_t \right)$, then the conditional variance of $x_t$ is proportional to its squared conditional mean, $\mu_t^2$. In the GLM literature, it is  the so called variance (or GLM) assumption. 
This implies that $x_t$ is conditionally heteroskedastic, thus enabling the model to capture the so called volatility of volatility (``vol-of-vol''). Nevertheless, according to the GLM assumption, the ratio between the conditional variance and the square of the conditional mean is constant. The latter, as stated below, has implications for estimating the asymptotic covariance matrix of the estimated parameters, while it is not required to get consistent estimates of the parameters entering in $\mu_t$; this is something that has not been emphasized in the MEM literature.

The nonnegativeness of the conditional mean could be ensured through the following sufficient constraints: $\omega>0$, $\alpha_p\geq0$, for $p=1,\dots,P$, $\beta_q\geq0$, for $q=1,\dots,Q$. Nevertheless, as noticed by \cite{Nelson:1992} in a GARCH framework, less severe constraints can be imposed to keep the conditional mean nonnegative. In the MEM($2$,$1$) case, which is of our interest, we impose the following constraints: $\omega>0$, $\alpha_1\geq0$, $\alpha_2+\alpha_1\beta_1\geq0$, and $0\leq\beta_1<1$. In other words, $\alpha_2$ can be negative.
Sufficient conditions for mean-stationarity of $x_t$ can be obtained by imposing $\sum_{p=1}^P\alpha_p + \sum_{q=1}^Q\beta_q<1 $. Then, the unconditional mean $x_t$ is equal to
\begin{equation}
	\E(x_t)= \mu= \dfrac{\omega}{1-\sum_{p=1}^P\alpha_p-\sum_{q=1}^Q\beta_q}. 
\end{equation}
The conditions for the existence of the unconditional variance are more involved; for the MEM(1,1), we have
\begin{equation}
	\Var(x_t)= \sigma^2= \mu^2\Var(\varepsilon_t)\dfrac{\left[1-(2\alpha_1\beta_1+\beta_1^2)\right]}{1-(\alpha_1+\beta_1)^2-\alpha_1^2\Var(\varepsilon_t)}=\mu^2\phi^2\dfrac{\left[1-(2\alpha_1\beta_1+\beta_1^2)\right]}{1-(\alpha_1+\beta_1)^2-\alpha_1^2\phi^2}. \label{eq:unc_var} 
\end{equation}

Hence, the unconditional second moment exists if $(\alpha_1+\beta_1)^2+\alpha_1^2\phi^2<1$. Furthermore, according to \eqref{eq:unc_var},  the ratio between the unconditional variance and the squared unconditional mean of $x_t$, i.e. its squared coefficient of variation, is greater than that of $\varepsilon_t$: the larger $\alpha_1$, the more dispersed $x_t$ than $\varepsilon_t$ \citep{Bauwens:2013}.\\
When $x_t$ refers to financial volatility measures, \eqref{eq:mean-MEM} could be augmented to capture the asymmetric response of the conditional variance to negative returns, the so called leverage effect. Then, the Asymmetric MEM, hereafter denoted by AMEM($P$,$Q$), proposed by \cite{Engle:Gallo:2006} specifies the conditional mean equation as follows
\begin{equation}
	\E(x_t|\mathcal{I}_{t-1})	=\mu_{t}= \omega + \sum_{p=1}^P\alpha_p x_{t-p}+ \sum_{q=1}^Q\beta_q \mu_{t-q} + \gamma_1x^{(-)}_{t-1},    
\end{equation}
where $x^{(-)}_{t}=x_t\mathbb{I}(r_{t}<0)$, $\mathbb{I}(\cdot)$ denotes the indicator function and $r_t$ is the corresponding financial return at time $t$. Under zero median of the returns \citep[see][]{Brownlees:2012}, mean-stationarity of $x_t$ implies 
\begin{equation}
	\E(x_t)= \mu= \dfrac{\omega}{1-\sum_{p=1}^P\alpha_p-\sum_{q=1}^Q\beta_q-\frac{\gamma_1}{2}}.
\end{equation}

\section{Model Estimation and Inference}\label{sec:est}

\subsection{Maximum Likelihood Estimation}\label{sec:mle}
Given a sample of $T$ observations, $x_1,\dots,x_T$, the log-likelihood function, conditional on the initial values, $\mu_0,\dots,\mu_{1-Q}$ and $x_0,\dots,x_{1-P}$, is the following
\begin{equation}
	\mathcal{L}(\bm{\theta})=\log f(x_1,\dots,x_T;\bm{\theta})=\sum_{t=1}^T\log f(x_t|\mathcal{I}_{t-1};\bm{\theta}),
\end{equation}
where $f(x_t|\mathcal{I}_{t-1})$ denotes the conditional probability density function of $x_t$, satisfying characteristics underlined in \eqref{errorMEM}, parameterized by a $k$ dimensional vector of parameters, $\bm{\theta}\in \bm{\Theta}\subseteq \mathbb{R}^k$. 
Within the MEM framework, it is customary to adopt the Gamma distribution for the error term in \eqref{errorMEM},\footnote{See \ref{gamma} for further details about the Gamma distribution and the respective probability density function.} given its nice inferential properties that will be discussed later. 

Hence, the Gamma log-likelihood is the following
\begin{equation}\label{eq:loglik-gamma-mem}
	\mathcal{L}(\bm{\theta})= \sum_{t=1}^Tl_t(\bm{\theta})= -\frac{1}{\phi^2}\sum_{t=1}^T\frac{x_t}{\mu_t}  + \frac{1}{\phi^2}\sum_{t=1}^T\ln \left(\dfrac{x_t}{\mu_t}\right) - \frac{T}{\phi^2}\ln \left(\phi^2\right) -\sum_{t=1}^T\ln\left(x_t\right) - T \ln\left[\Gamma\left(\frac{1}{\phi^2}\right)\right],
\end{equation}
with $\Gamma\left(z\right)$ the Gamma function, while $\mu_t$ can be computed recursively through \eqref{condvar}. Given its recursive structure, initial values of the conditional mean and $x_t$, namely $\mu_0,\dots,\mu_{1-Q}$ and $x_0,\dots,x_{1-P}$, are required. A suitable choice is to employ an estimator of their unconditional expectation. Following standard practice, we rely on the sample average of $x_t$, that is $\mu_0=\dots=\mu_{1-Q}=x_0=\dots=x_{1-P}=T^{-1}\sum_{t=1}^Tx_t$.\\
We denote by  $l_t(\bm{\theta})=\log f(x_t|\mathcal{I}_{t-1};\bm{\theta})$ the contribution to the log-likelihood of the $t$th observation, while $\bm{\lambda}=(\omega,\alpha_1,\dots,\alpha_p,\beta_1,\dots,\beta_q,\gamma_1)'$ is the vector collecting the conditional mean equation parameters, so we can partition the vector of all parameters as follows, $\bm{\theta}=(\bm{\lambda}',\phi^2)'$.\footnote{We are considering the most general case, that is, the AMEM($P$,$Q$).}\\
\cite{Fiorentini:1996} computed first- and second-order derivatives of the GARCH log-likelihood, obtaining a recursive structure. Hereafter, we show the analytical derivatives of the MEM log-likelihood, up to the second order, while keeping the notation close to \cite{Fiorentini:1996}, when the Gamma distribution is assumed for the error term.
The first-derivative vector of the log-likelihood function, also referred to as the score, is $\frac{\partial \mathcal{L}(\bm{\theta})}{\partial\bm{\theta}}=\sum_{t=1}^T\frac{\partial l_t(\bm{\theta})}{\partial\bm{\theta}}$, with $\frac{\partial l_t(\bm{\theta})}{\partial\bm{\theta}}$  the contribution to the score of the $t$th observation, whose elements are given in \eqref{eq:foc_mu_gamma-MEM} and \eqref{eq:foc_phi_gamma-MEM}.
\begin{equation}
	\frac{\partial l_t(\bm{\theta})}{\partial\bm{\lambda}}=  \frac{1}{\phi^2}\frac{x_t-\mu_t}{\mu_t^2}\frac{\partial\mu_t}{\partial\bm{\lambda}}=\frac{1}{\phi^2}\left(\frac{x_t}{\mu_t}-1\right)\frac{1}{\mu_t}\frac{\partial\mu_t}{\partial\bm{\lambda}}, \label{eq:foc_mu_gamma-MEM}
\end{equation}
where 
\begin{equation*}
	\frac{\partial\mu_t}{\partial\bm{\lambda}}=\bm{z}_t+\sum_{q=1}^Q\beta_q\frac{\partial\mu_{t-q}}{\partial\bm{\lambda}},
\end{equation*}
with $\bm{z}_t=(1,x_{t-1},\dots,x_{t-p},\mu_{t-1},\dots,\mu_{t-q},x^{(-)}_{t-1})'$.  The initialization of the conditional mean, $\mu_t$, discussed above, implies that for $t\leq0$
\begin{equation}
	\frac{\partial\mu_t}{\partial\bm{\lambda}}=\bm{0},
\end{equation}
since it does not depend on the vector of parameters.
Conversely, the contribution to the score of the $t$th observation with respect to $\phi^2$ is given by

\begin{equation}
	\frac{\partial l_t(\bm{\theta})}{\partial\phi^2}= \frac{1}{\phi^4} \left[\frac{x_t}{\mu_t}-\ln \left(\dfrac{x_t}{\mu_t}\right)+\bm{\Psi}\left(\frac{1}{\phi^2}\right)+\ln\left(\phi^2\right)-1\right],  \label{eq:foc_phi_gamma-MEM}
\end{equation}
with $\bm{\Psi}\left(z\right)=\frac{\partial}{\partial z}\ln\Gamma\left(z\right)$ the digamma function.\\
The Hessian matrix, $\bm{H}(\bm{\theta})=\sum_{t=1}^T\bm{H}_t(\bm{\theta})$, the matrix of second-order derivatives, where $\bm{H}_t(\bm{\theta})=\frac{\partial^2 l_t(\bm{\theta})}{\partial\bm{\theta}\partial\bm{\theta}'}$,	can be partitioned as follows
\begin{equation*}
	\bm{H}_t(\bm{\theta})= \begin{bmatrix}
		\frac{\partial^2 l_t(\bm{\theta})}{\partial\bm{\lambda}\partial\bm{\lambda}'} & 	\frac{\partial^2 l_t(\bm{\theta})}{\partial\bm{\lambda}\partial\phi^2} \\
		\frac{\partial^2 l_t(\bm{\theta})}{\partial\phi^2\partial\bm{\lambda}'}	& \frac{\partial^2 l_t(\bm{\theta})}{\partial\left(\phi^2\right)^2}
	\end{bmatrix}.
\end{equation*}
The blocks are given by the following equations
\begin{align}
	\frac{\partial^2l_t(\bm{\theta})}{\partial\bm{\lambda}\partial\bm{\lambda}'}&=	 \frac{1}{\phi^2}\left(\frac{x_t}{\mu_t}-1\right)\left(-\frac{1}{\mu_t^2}\frac{\partial\mu_t}{\partial\bm{\lambda}}\frac{\partial\mu_t}{\partial\bm{\lambda}'} + \frac{1}{\mu_t}\frac{\partial^2\mu_t}{\partial\bm{\lambda}\partial\bm{\lambda}'}\right)-\frac{1}{\phi^2} \frac{x_t}{\mu_t}\frac{1}{\mu^2_t}\frac{\partial\mu_t}{\partial\bm{\lambda}}\frac{\partial\mu_t}{\partial\bm{\lambda}'},  \label{eq:hess_mu_gamma-MEM} \\
	\frac{\partial^2l_t(\bm{\theta})}{\partial\left(\phi^2\right)^2}&=  \frac{1}{\phi^6}\left[ 3 -2\bm{\Psi}\left(\frac{1}{\phi^2}\right)-\frac{1}{\phi^2}\bm{\Psi}'\left(\frac{1}{\phi^2}\right) -2\ln\left(\phi^2\right)+2\ln\left( \frac{x_t}{\mu_t}\right)-2\frac{x_t}{\mu_t}\right], \label{eq:hess_phi_gamma-MEM} \\
	\frac{\partial^2l_t(\bm{\theta})}{\partial\bm{\lambda}  \partial\phi^2}&=	-\frac{1}{\phi^4}\frac{x_t-\mu_t}{\mu_t^2}\frac{\partial\mu_t}{\partial\bm{\lambda}}=-\frac{1}{\phi^4}\left(\frac{x_t}{\mu_t}-1\right)\frac{1}{\mu_t}\frac{\partial\mu_t}{\partial\bm{\lambda}},  \label{eq:hess_phi_mu_gamma-MEM}    
\end{align}
with $\bm{\Psi}'\left(z\right)=\frac{\partial^2}{\partial z^2}\ln\Gamma\left(z\right)=\frac{\partial}{\partial z}\bm{\Psi}\left(z\right)$ the trigamma function.\\
The second-order derivatives of the conditional mean equation are specified as follows
\begin{equation}
	\frac{\partial\mu^2_t}{\partial\bm{\lambda}\partial\bm{\lambda}'}= \frac{\partial 	\bm{z}_t}{\partial\bm{\lambda}'} +\sum_{q=1}^Q\frac{\partial\mu_{t-q}}{\partial\bm{\lambda}}\frac{\partial \beta_q}{\partial\bm{\lambda}'} + \sum_{q=1}^Q\beta_q \frac{\partial\mu^2_{t-q}}{\partial\bm{\lambda}\partial\bm{\lambda}'}, 
\end{equation}
where
\begin{equation}
	\frac{\partial \bm{z}_t}{\partial\bm{\lambda}'}= \begin{bmatrix}
		0 & 0 &\cdots & 0 & 0  &\cdots & 0 &0\\
		\vdots & \vdots & & \vdots& \vdots  &  & \vdots& \vdots\\
		0	& 0 &\cdots & 0 & 0 &\cdots & 0 & 0\\
		\frac{\partial\mu_{t-1}}{\partial{\omega}}&\frac{\partial\mu_{t-1}}{\partial{\alpha_1}} & \ldots& \frac{\partial\mu_{t-1}}{\partial{\alpha_p}}  &\frac{\partial\mu_{t-1}}{\partial{\beta_1}} & \ldots & \frac{\partial\mu_{t-1}}{\partial{\beta_q}} & \frac{\partial\mu_{t-1}}{\partial{\gamma_1}} \\
		\vdots &\vdots  & & \vdots & \vdots & & \vdots & \vdots  \\
		\frac{\partial\mu_{t-q}}{\partial{\omega}}&\frac{\partial\mu_{t-q}}{\partial{\alpha_1}} & \ldots& \frac{\partial\mu_{t-q}}{\partial{\alpha_p}}  &\frac{\partial\mu_{t-q}}{\partial{\beta_1}} & \ldots & \frac{\partial\mu_{t-q}}{\partial{\beta_q}} & \frac{\partial\mu_{t-q}}{\partial{\gamma_1}} \\
	\end{bmatrix},  
\end{equation}
with  $\left(\sum_{q=1}^Q\frac{\partial\mu_{t-q}}{\partial\bm{\lambda}}\frac{\partial \beta_q}{\partial\bm{\lambda'}}\right)'=\frac{\partial \bm{z}_t}{\partial\bm{\lambda'}}$, and $\frac{\partial \beta_q}{\partial\bm{\lambda'}}$ is a $(P+Q+2)$ vector whose $(1+P+q)$th element is equal to one, i.e., $\frac{\partial \beta_q}{\partial\bm{\lambda'}}=(0,\dots,0,1,\dots,0)$.\\
The MLE of $\bm{\theta}$, $\hat{\bm{\theta}}$ is given by
\begin{equation}
	\hat{\bm{\theta}}= \argmax_{\bm{\theta} \in \bm{\Theta}} \mathcal{L}(\bm{\theta}),
\end{equation}
which is the root of the following likelihood equations that globally maximizes the log-likelihood
\begin{equation}\label{foc}
	\begin{split}
		\frac{\partial \mathcal{L}(\bm{\theta})}{\partial\bm{\lambda}}&=  \frac{1}{\phi^2}\sum_{t=1}^T\frac{x_t-\mu_t}{\mu_t^2}\frac{\partial\mu_t}{\partial\bm{\lambda}}=\frac{1}{\phi^2}\sum_{t=1}^T\left(\frac{x_t}{\mu_t}-1\right)\frac{1}{\mu_t}\frac{\partial\mu_t}{\partial\bm{\lambda}}=0, \\
		\frac{\partial \mathcal{L}(\bm{\theta})}{\partial\phi^2}&= \frac{1}{\phi^4} \sum_{t=1}^T\left[\frac{x_t}{\mu_t}-\ln \left(\dfrac{x_t}{\mu_t}\right)+\bm{\Psi}\left(\frac{1}{\phi^2}\right)+\ln\left(\phi^2\right)-1\right]=0.  
	\end{split}   
\end{equation}
Nevertheless, the likelihood equations are nonlinear functions of the model parameters that cannot be solved analytically and a numerical maximization procedure is needed. The MLE of the model parameters is then computed via some iterative algorithms, one of which will be described in the next subsection. \\
Under standard regularity conditions,\footnote{See, for example, \cite{Engle:Russell:1998}.} $\hat{\bm{\theta}}$ is consistent and its asymptotic distribution is given by\footnote{Hereafter, we use the same symbol $\bm{\theta}$ to denote both a generic parameter value and the true parameter value, given that the distinction could be rather cumbersome.}
\begin{equation}
	\sqrt{T}(\hat{\bm{\theta}}-\bm{\theta}) \stackrel{d}{\rightarrow} \mathcal{N}(\bm{0},\bm{A}^{-1}),
\end{equation}\label{eq:as_dist_MLE}
with 
\begin{equation}
	\bm{A}=-\E\left[\bm{H}_t(\bm{\theta})\right],
\end{equation}
the expectation of the Hessian matrix of $l(\bm{\theta}_t)$. 
By the law of iterated expectations, $\E\left[\bm{H}_t(\bm{\theta})\right]= \E\lbrace\E\left[\bm{H}_t(\bm{\theta})|\mathcal{I}_{t-1}\right]\rbrace$, whose elements are given below
\begin{align}
	\E	\left[\frac{\partial^2 l_t(\bm{\theta})}{\partial\bm{\lambda}\partial\bm{\lambda}'}|\mathcal{I}_{t-1}\right]&= 	- \frac{1}{\phi^2}\frac{1}{\mu_t^2}\frac{\partial\mu_t}{\partial\bm{\lambda}}\frac{\partial\mu_t}{\partial\bm{\lambda}'}, \label{eq:exp_hess_mu_gamma-MEM} \\
	\E	\left[\frac{\partial^2 l_t(\bm{\theta})}{\partial\left(\phi^2\right)^2}|\mathcal{I}_{t-1}\right]&=  \frac{1}{\phi^6}\left\{ 3 -2\bm{\Psi}\left(\frac{1}{\phi^2}\right)-\frac{1}{\phi^2}\bm{\Psi}'\left(\frac{1}{\phi^2}\right) -2\ln\left(\phi^2\right)+2\E\left[\ln\left( \epsilon_{t}\right)\right]-2\right\} \label{eq:exp_hess_phi_gamma-MEM} \\ 
	&= \frac{1}{\phi^6}\left[ 3 -2\bm{\Psi}\left(\frac{1}{\phi^2}\right)-\frac{1}{\phi^2}\bm{\Psi}'\left(\frac{1}{\phi^2}\right) -2\ln\left(\phi^2\right)+2\bm{\Psi}\left(\frac{1}{\phi^2}\right)+2\ln\left(\phi^2\right)-2\right]  \nonumber \\
	&= \frac{1}{\phi^6}\left[ 1 -\frac{1}{\phi^2}\bm{\Psi}'\left(\frac{1}{\phi^2}\right) \right] \nonumber, \\
	\E	\left[\frac{\partial^2 l_t(\bm{\theta})}{\partial\bm{\lambda}\partial\phi^2}|\mathcal{I}_{t-1}\right]&=\bm{0}, \label{eq:exp_hess_phi_mu_gamma-MEM} 
\end{align}
where the equality from the first line to the second
line in \eqref{eq:exp_hess_phi_gamma-MEM}, follows from $\E\left[\ln(\varepsilon_{t})\right] = \bm{\Psi}\left(\frac{1}{\phi^2}\right) + \ln(\phi^2)$.\footnote{See  \cite{Park:2009}.}
As stated in \eqref{eq:exp_hess_phi_mu_gamma-MEM}, the information matrix (IM) is block diagonal\footnote{This implies that the ML estimators of $\bm{\lambda}$ and $\phi^2$ are asymptotically uncorrelated.}
\begin{equation}\label{eq:A-ML}
	\bm{A}= \begin{bmatrix}
		\bm{A}^{\bm{\lambda}} & 0 \\
		0& \bm{A}^{\phi^2},  
	\end{bmatrix},
\end{equation}

$\E	\left[\frac{\partial^2 l_t(\bm{\theta})}{\partial\left(\phi^2\right)^2}\right]$ is nonstochastic, while $\E\left[\frac{\partial^2 l_t(\bm{\theta})}{\partial\bm{\lambda}\partial\bm{\lambda}'}\right]$ can be estimated through its in-sample counterpart, then
\begin{equation}
	\hat{\bm{A}}^{\bm{\lambda}}_{IM}= T^{-1}\sum_{t=1}^T\frac{1}{\hat{\phi}^2}\frac{1}{\hat{\mu_t}^2}\frac{\partial\mu_t}{\partial\hat{\bm{\lambda}}}\frac{\partial{\mu}_t}{\partial\hat{\bm{\lambda}}'},
\end{equation}    
where $\hat{\mu_t}$ denotes ${\mu_t}$ evaluated at $\hat{\bm{\theta}}$.
An alternative estimator of $\bm{A}$ is based on the observed Hessian (H) matrix,
\begin{equation}\label{eq:actual_hessina}
	\hat{\bm{A}}_H= -T^{-1}\sum_{t=1}^T\frac{\partial^2 l_t(\bm{\theta})}{\partial\hat{\bm{\theta}}\partial\hat{\bm{\theta}}'},
\end{equation}   
whose elements can be retrieved from \eqref{eq:hess_mu_gamma-MEM}-\eqref{eq:hess_phi_mu_gamma-MEM}. It is important to stress that the first estimator of $\bm{A}$, $\hat{\bm{A}}_{IM}$,  depends only on the first derivatives of the conditional mean equation, while the second requires second-order derivatives.
Another estimator of $\bm{A}$ is based on the outer product of gradients (OPG), which relies on first-order derivatives of the log-likelihood
\begin{equation}\label{eq:actual_opg}
	\hat{\bm{A}}_{OPG}= T^{-1}\sum_{t=1}^T\frac{\partial l_t(\bm{\theta})}{\partial\hat{\bm{\theta}}}\frac{\partial l_t(\bm{\theta})}{\partial\hat{\bm{\theta}}'}.
\end{equation}
Nevertheless, it is widely accepted in the econometric literature that it performs poorly in small samples, especially in dynamic models \citep{Davidson:1993}.
\subsection{The Scoring Algorithm}\label{sec:scoring}
To compute the MLE, an iterative procedure is required, since a closed-form solution for the likelihood equations in \eqref{foc} is not available. Hereafter, we rely on the scoring algorithm, an iterative algorithm for finding the roots of the likelihood equations, belonging to the gradient method family (more specifically the modified Newton's one), which is specified as follows 
\begin{equation}\label{eq:scoring}
	\ddot{\bm{\theta}}=\dot{\bm{\theta}} -\left\lbrace\E\left[{\bm{H}(\dot{\bm{\theta}})}\right]\right\rbrace^{-1}s(\dot{\bm{\theta}}).
\end{equation}
Parameters and quantities specified as functions of the parameters labelled with one dot denote updates from the previous iteration, whereas those marked with two dots signify updates from the current iteration.
Then, $\ddot{\bm{\theta}}$ is the update of the vector of $k$ unknown parameters at the current iteration, $s(\dot{\bm{\theta}})$ is the score evaluated at $    \dot{\bm{\theta}}$, while $\E\left[{\bm{H}(\dot{\bm{\theta}}})\right]$ is the update of the expected value of the Hessian. Given an initial guess for $\bm{\theta}$, we can iterate \eqref{eq:scoring} until the difference among two consecutive log-likelihoods (evaluated at $\ddot{\bm{\theta}}$ and $\dot{\bm{\theta}}$) is below a prespecified threshold (e.g., $1e\text{-}8$).
As stated below, the method of scoring is useful when the expected value of the Hessian matrix is block diagonal, since the updating of the parameters of each block can be computed separately. Furthermore, if the observed Hessian matrix fails to be negative definite — which is likely when $\dot{\bm{\theta}}$ is far from the maximum — the optimization algorithm may diverge. In contrast, for our purposes, the expected Hessian matrix is negative definite by construction (see \eqref{eq:exp_hess_mu_gamma-MEM} and \eqref{eq:exp_hess_phi_gamma-MEM}), which avoids one potential source of instability in the optimization procedure.

Then, the formulae in \eqref{eq:foc_mu_gamma-MEM}-\eqref{eq:foc_phi_gamma-MEM} and \eqref{eq:exp_hess_mu_gamma-MEM}-\eqref{eq:exp_hess_phi_mu_gamma-MEM} can be inserted in \eqref{eq:scoring}, thus obtaining
\begin{align}
	\ddot{\bm{\lambda}}&=\dot{\bm{\lambda}} + \left[\sum_{t=1}^T\frac{1}{\dot{\phi}^2\dot{\mu}_t^2}\frac{\partial\mu_t}{\partial\dot{\bm{\lambda}}}\frac{\partial\mu_t}{\partial\dot{\bm{\lambda}}'}\right]^{-1} \sum_{t=1}^T\frac{1}{\dot{\phi}^2}\left(\frac{x_t}{\dot{\mu}_t}-1\right)\frac{1}{\dot{\mu_t}}\frac{\partial\mu_t}{\partial\dot{\bm{\lambda}}}\label{eq:scoring_mu-MEM}\\ 
	&=\dot{\bm{\lambda}}+\left[\dot{\bm{C}}'\dot{\bm{C}}\right]^{-1}\dot{\bm{C}}'\dot{\bm{z}}\nonumber\\ 
	\ddot{\phi}^2&=\dot{\phi}^2 + \frac{\dot{\phi}^6}{T}\left[\frac{\dot{\phi}^2}{-\dot{\phi}^2+\bm{\Psi}'\left(\frac{1}{\dot{\phi}^2}\right)}\right]\frac{1}{\dot{\phi}^4} \left[\sum_{t=1}^T\frac{x_t}{\dot{\mu}_t}-\sum_{t=1}^T\ln \left(\frac{x_t}{\dot{\mu}_t}\right)+T\bm{\Psi}\left(\frac{1}{\dot{\phi}^2}\right)+T\ln\left(\dot{\phi}^2\right)-T\right], \label{eq:scoring_phi-MEM} \\
	&=\dot{\phi}^2 + \left[\frac{\dot{\phi}^4}{-\dot{\phi}^2+\bm{\Psi}'\left(\frac{1}{\dot{\phi}^2}\right)}\right]\left[T^{-1}\sum_{t=1}^T\frac{x_t}{\dot{\mu}_t}-T^{-1}\sum_{t=1}^T\ln\left(\frac{x_t}{\dot{\mu}_t}\right)+\bm{\Psi}\left(\frac{1}{\dot{\phi}^2}\right)+\ln\left(\dot{\phi}^2\right)-1\right] \nonumber \\
	&=\dot{\phi}^2 + \left[\frac{\dot{\phi}^4}{-\dot{\phi}^2+\bm{\Psi}'\left(\frac{1}{\dot{\phi}^2}\right)}\right]\left[T^{-1}\sum_{t=1}^T\left(\frac{x_t}{\dot{\mu}_t}-1\right)-T^{-1}\sum_{t=1}^T\ln\left(\frac{x_t}{\dot{\mu}_t}\right)+\bm{\Psi}\left(\frac{1}{\dot{\phi}^2}\right)+\ln\left(\dot{\phi}^2\right)\right] \nonumber \\
	&=\dot{\phi}^2 + \left[\frac{\dot{\phi}^4}{-\dot{\phi}^2+\bm{\Psi}'\left(\frac{1}{\dot{\phi}^2}\right)}\right]\left[T^{-1}\sum_{t=1}^T\dot{v}_t-T^{-1}\sum_{t=1}^T\ln\left(\frac{x_t}{\dot{\mu}_t}\right)+\bm{\Psi}\left(\frac{1}{\dot{\phi}^2}\right)+\ln\left(\dot{\phi}^2\right)\right], \nonumber 
\end{align}
where the $t$th row of $\bm{C}$, $\bm{c}'_t$, and the $t$th element of the column vector $\bm{z}$ are defined below $$\bm{c}_t=\left(\frac{1}{\phi\mu_t}\frac{\partial\mu_t}{\partial\bm{\lambda}}\right), \quad z_t=\frac{1}{\phi}\left(\frac{x_t}{\mu_t}-1\right)=\frac{v_t}{\phi},$$
with $z_t$ the standardized error (zero mean and unit variance), while $v_t$ is the zero-mean error, with variance equal to $\phi^2$.
Then, $\ddot{\bm{\lambda}}$ is obtained by adding the coefficient vectors in the least square regression of $\dot{z}_t$ on $\dot{\bm{c}}_t$, to $\dot{\bm{\lambda}}$.
Note also that, at convergence, the OLS slope vector is equal to 0, this happens when $\frac{\partial \mathcal{L}(\bm{\theta})}{\partial\bm{\lambda}}=\bm{0}$, that is, $\dot{\bm{C}}'\dot{\bm{z}}=0$. \\     
The scoring algorithm adopted in \eqref{eq:scoring_mu-MEM} to obtain the MLE of $\bm{\lambda}$ can be reformulated as an Iterative Reweighted Least Squares (IRLS), where the observations are reweighted at each iteration and the update follows a weighted regression scheme
\begin{align}
	\ddot{\bm{\lambda}}&=\dot{\bm{\lambda}} + \left[\sum_{t=1}^T\frac{1}{\dot{\mu}_t^2}\frac{\partial\mu_t}{\partial\dot{\bm{\lambda}}}\frac{\partial\mu_t}{\partial\dot{\bm{\lambda}}'}\right]^{-1} \sum_{t=1}^T\frac{1}{\dot{\mu}_t}\left(\frac{x_t}{\dot{\mu}_t}-1\right)\frac{\partial\mu_t}{\partial\dot{\bm{\lambda}}}=\dot{\bm{\lambda}}+\left(\dot{\bm{C}}'\dot{\bm{C}}\right)^{-1}\dot{\bm{C}}'\dot{\bm{v}} \label{eq:IRLS_mu-MEM}\\ 
	&=  \left(\dot{\bm{C}}_{*}'\dot{\bm{W}}\dot{\bm{C}}_{*}\right)^{-1}\dot{\bm{C}}_{*}'\dot{\bm{W}}\dot{\bm{v}}_{*} \nonumber
\end{align}
where $\bm{W}=\text{diag}\left(\bm{w}\right)$. The $t$th row of $\bm{C}_{*}$, $\bm{c}'_{t*}$  and the $t$th element of the column vectors $\bm{v}_{*}$ and $\bm{w}$ are defined below $$\bm{c}_{t*}=\left(\frac{\partial\mu_t}{\partial\bm{\lambda}}\right), \quad v_{t*}=\left(\frac{\partial\mu_t}{\partial\bm{\lambda}'}\bm{\lambda}+\frac{x_t}{\mu_t}-1\right), \quad w_t=\left(\frac{1}{\phi^2\mu^2_t}\right).$$ 
Then, the update of ${\bm{\lambda}}$, namely $\ddot{\bm{\lambda}}$ is computed by regressing $\dot{v_{t*}}$ on $\dot{\bm{c}}_{t*}$, with weights $\dot{w}_{t}$.\\

Finally, at convergence, notice that the estimated asymptotic variance of $\hat{\bm{\lambda}}$ through the information matrix, is equal to $\left(\hat{\bm{C}}'\hat{\bm{C}}\right)^{-1}$, that is, the estimated covariance matrix of $\hat{\bm{\lambda}}$ is the same as the quantity used in the last iteration of the scoring algorithm.

\subsection{Quasi Maximum Likelihood Estimator}\label{sec:qmle}

In principle, any distribution for $\varepsilon_{t}$ satisfying the properties indicated in the previous section, i.e., positive support and unit--mean expectation, could be adopted. Nevertheless, notice that the first-order conditions in \eqref{eq:foc_mu_gamma-MEM}, based on the Gamma log-likelihood, do not depend on the parameter $\phi^2$, and, if the conditional mean equation, $\mu_t$, is correctly specified, the expected value of the score is zero when evaluated at the true parameter value. Then, even though the actual conditional distribution of $x_t$ is not Gamma, the maximization of the Gamma log-likelihood function (called quasi log-likelihood) provides consistent estimates of the conditional mean equation parameters, $\bm{\lambda}$  \citep[see, e.g., ][]{Engle:Gallo:2006}. The estimator maximizing the misspecified log-likelihood function is referred to as the  Quasi Maximum Likelihood Estimator (QMLE),\footnote{See, e.g., for the properties of QMLE in static models, \cite{White82,GourierouxQMLE84}, and \cite{Bollerslev92} for dynamic models.} given that the density used in the log-likelihood is different from the actual density of $x_t$. 
The asymptotic distribution of the QMLE of $\bm{\lambda}$, under suitable regularity conditions,\footnote{See, once again, \cite{Engle:Russell:1998}.} is given by
\begin{equation}
	\sqrt{T}(\hat{\bm{\lambda}}-\bm{\lambda}) \stackrel{d}{\rightarrow} \mathcal{N}(\bm{0},\bm{A}^{\bm{\lambda}^{-1}}\bm{B}^{\bm{\lambda}}\bm{A}^{\bm{\lambda}^{-1}})
\end{equation}
where $\bm{A}^{\bm{\lambda}}$ has been previously defined in \eqref{eq:A-ML}, while $\bm{B}^{\bm{\lambda}}$
\begin{equation}
	\bm{B}^{\bm{\lambda}}=	\E\left(\frac{\partial l_t(\bm{\lambda})}{\partial\bm{\lambda}}\frac{\partial l_t(\bm{\lambda})}{\partial\bm{\lambda}}'\right),    
\end{equation}
is the covariance of $s_t(\bm{\lambda})$. $\E\left[S(\bm{\lambda})S(\bm{\lambda})'\right]=\sum_{t=1}^T\E\left[s_t(\bm{\lambda})s_t(\bm{\lambda})'\right]$,\footnote{Note that the scores are serially uncorrelated.} where $\E\left[s_t(\bm{\lambda})s_t(\bm{\lambda})'\right]$ is given by
\begin{align}
	\begin{split}
		\E\left[s_t(\bm{\lambda})s_t(\bm{\lambda}')\right]	&=\E\left[\frac{1}{\phi^4}\left(\frac{x_t}{\mu_t}-1\right)^2\frac{1}{\mu^2_t}\frac{\partial\mu_t}{\partial\bm{\lambda}}\frac{\partial\mu_t}{\partial\bm{\lambda}'}\right]=\frac{1}{\phi^4}\E\left[\frac{v_t^2}{\mu^2_t}\frac{\partial\mu_t}{\partial\bm{\lambda}}\frac{\partial\mu_t}{\partial\bm{\lambda}'}\right].\\  \nonumber
		\label{eq:exp_OPG_mu_gamma-MEM}
	\end{split}
\end{align}
Then, a consistent estimator of ${\Var}(\hat{\bm{\lambda}})$, referred to as fully robust (FR), is
\begin{equation}\label{eq:robust}
	\widehat{\Var}_{FR}(\hat{\bm{\lambda}})=T^{-1}\hat{\bm{A}}^{\bm{\lambda}^{-1}}_{IM}\hat{\bm{B}}_{OPG}^{\bm{\lambda}}\hat{\bm{A}}^{\bm{\lambda}^{-1}}_{IM}=\left[\sum_{t=1}^T\frac{1}{\hat{\mu}_t^2}\frac{\partial \mu_t}{\partial \hat{\bm{\lambda}}}\frac{\partial \mu_t}{\partial \hat{\bm{\lambda}}'}\right]^{-1}\sum_{t=1}^T\frac{\hat{v}_t^2}{\hat{\mu}^2_t}\frac{\partial\mu_t}{\partial\hat{\bm{\lambda}}}\frac{\partial\mu_t}{\partial\hat{\bm{\lambda}}'}
	\left[\sum_{t=1}^T\frac{1}{\hat{\mu}_t^2}\frac{\partial \mu_t}{\partial \hat{\bm{\theta}}}\frac{\partial \mu_t}{\partial \hat{\bm{\theta}}'} \right]^{-1} .
\end{equation}
As noticed by \cite{Engle:Gallo:2006}, the $\phi^2$'s cancel out, thus their values do not influence the estimation of the QMLE covariance matrix.
More specifically, the estimator suggested by \cite{Engle:Gallo:2006}, usually referred to as the sandwich (S) estimator, approximates the matrix $\bm{A}$ through the actual Hessian (see \eqref{eq:actual_hessina}),
which requires the computation of second-order derivatives, while the estimator described in \eqref{eq:robust} depends only on first-order derivatives of the conditional mean equation, and when the inverse exists, the covariance matrix is always nonnegative definite. Nevertheless, both estimators, S and FR, do not require any variance assumption and are consistent under general forms of heteroskedasticity (we will clarify this point later).

Nevertheless, the MEM is usually based on the following conditional variance assumption
\begin{equation}\label{GLM}
	\Var(x_t|\mathcal{I}_{t-1})=\phi^2\mu^2_{t},
\end{equation} see, e.g., \cite{Brownlees:2012}. In other words, they postulate a constant ratio between the conditional variance and the squared conditional mean; this assumption is also referred to as the GLM or variance assumption \citep{Wooldridge:1999}. As a consequence, by noticing that
$\phi^2=E\left(\dfrac{u^2_t}{\mu_t^2}\right)$ by the law of iterated expectation, a consistent method of moments estimator of $\phi^2$ is the following
\begin{equation}\label{est_sigma}
	\hat\phi^2=T^{-1}\sum_{t=1}^T\dfrac{\hat u_t^2}{\hat\mu_t^2},
\end{equation}
see also \cite{Brownlees:2012}. Then, if the actual conditional distribution of the data is not Gamma, the MLE of $\phi^2$ is not consistent \citep[see, e.g.,][]{Mccullagh:2019}, and it is also unfeasible if there are some zeros in the data. Whereas the method of the moments estimator in \eqref{est_sigma}, requires weaker assumptions than the MLE does (see also the first-order conditions in \eqref{eq:foc_phi_gamma-MEM}). 
Under the GLM assumption $\bm{B}_{\lambda}=\bm{A}_{\bm{\lambda}}$, and the estimator of $\Var(\hat{\bm{\lambda}})$ reduces to
\begin{equation}\label{eq:glm}
	\widehat{\Var}_{GLM}(\hat{\bm{\lambda}})= 
	\hat{\phi}^2\left(\sum_{t=1}^T\frac{1}{\hat{\mu}_t^2}\frac{\partial \mu_t}{\partial \hat{\bm{\lambda}}}\frac{\partial \mu_t}{\partial \hat{\bm{\lambda}}'} \right)^{-1} .
\end{equation}

\begin{remark}
	The QMLE of the conditional mean equation parameters, $\bm{\lambda}$, is consistent if $\E(x_t|\mathcal{I}_{t-1})=\mu_t$. Consistency does not require either a particular distributional assumption or the GLM assumption, $\Var(x_t|\mathcal{I}_{t-1})=\phi^2\mu^2_t$. 
\end{remark}
\begin{remark}
	The conditional variance assumption adopted in \eqref{eq:glm} is needed for the estimation of the variance-covariance matrix of the QMLE of ${\bm{\lambda}}$. Obviously, it is just a moment condition and does not imply that the conditional distribution of the data is Gamma.
\end{remark}

\begin{remark}
	The fully robust estimator of the variance-covariance matrix of $\bm{\lambda}$ QMLE proposed in \eqref{eq:robust} does not require the GLM assumption, $\Var(x_t|\mathcal{I}_{t-1})=\phi^2\mu^2_t$ to be consistent. 
\end{remark}

\section{Simulation Study}\label{sim}
In this Section, we evaluate the finite-sample behaviour of several estimators of the QMLE asymptotic covariance matrix through an extensive simulation study.\footnote{Note that the MLE is a special case of the QMLE.}  We assume that the series are generated from an AMEM(1,1), as described in \eqref{errorMEM}-\eqref{condvar},\footnote{To speed up the estimation procedure, we do not consider further lags of the dependent variable, but generalization to higher-order lags is straightforward.} characterized by the vector of conditional mean equation parameters whose value is equal to $(\omega,\alpha_1,\beta_1,\gamma_1)'=(1.500,0.400,0.500,0.050)'$  and the dispersion parameter $\phi^2=0.0625$. The returns (required to simulate the sign for the
asymmetric effects) are generated from an AR(1) process with an unconditional mean of $0.02$, an autoregressive coefficient equal to $-0.02$, and standard Gaussian errors. We set the number of replications ($R$) to 10000 across sample sizes  $T\in\left\lbrace500,1000,2500,5000\right\rbrace$. For each replication, $r=1,\dots,R$, we compute the MLEs of the model parameters using the scoring algorithm illustrated in Section \ref{sec:scoring}, along with the corresponding standard errors derived from six covariance matrix estimators presented in Section \ref{sec:qmle}, namely: Hessian (H), outer product of gradients (OPG), information matrix (IM), sandwich (S), GLM, and fully robust (FR). \\
Let $\hat{\bm{\lambda}}^{(r)}$ denote the MLE of $\bm{\lambda}$ obtained in the $r$th replication, the approximation to the true standard error of $\hat{{\lambda}}_k$, where the latter denotes the $k$th element of $\hat{\bm{\lambda}}$, is computed as follows\footnote{Notice that the true value of the standard error is not available analytically in our estimation framework.}
\begin{equation}
	\text{SE}(\hat{\lambda}_k)=\sqrt{R^{-1}\sum_{r=1}^R\left(\hat{\lambda}_k^{(r)}-\bar{\lambda}_k\right)^2}, \qquad  \bar{\lambda}_k=R^{-1}\sum_{r=1}^R\hat{\lambda}_{k}^{(r)}.
\end{equation}
Moreover, we also evaluate the closeness of the estimated standard errors to the true standard errors. To this end, we compute the root mean squared error (RMSE) of the different approaches adopted to estimate standard errors.\footnote{For further details, see the simulation setup in \cite{Boldea:2009}.}
Then, for each approach the following quantities are calculated
\begin{equation}
	\text{S}_{1k}= R^{-1}\sum_{r=1}^R\widehat{\text{SE}}(\hat{\lambda}_k^{(r)}), \qquad  \text{S}_{2k}= R^{-1}\sum_{r=1}^R\left[\widehat{\text{SE}}(\hat{\lambda}_k^{(r)})\right]^2, \qquad \text{V}\left[\widehat{\text{SE}}(\hat{\lambda}_k)\right]= \text{S}_{2k} - \text{S}^2_{1k},
\end{equation}
where $\widehat{\text{SE}}(\hat{\lambda}_k)$ are the estimated standard errors of $\lambda_k$ for each covariance estimator, whose bias and RMSE are obtained as follows
\begin{equation}
	\text{BIAS}\left[\widehat{\text{SE}}(\hat{\lambda}_k)\right]= \text{S}_{1k}-  \text{SE}(\hat{\lambda}_k), \qquad \text{RMSE}\left[\widehat{\text{SE}}(\hat{\lambda}_k)\right]=\sqrt{\text{V}\left[\widehat{\text{SE}}(\hat{\lambda}_k)\right]+\text{BIAS}^2\left[\widehat{\text{SE}}(\hat{\lambda}_k)\right]}.
\end{equation}
Overall, we consider five simulation cases, grouped into four experimental settings
\begin{enumerate}
	\item \textit{Correct specification} (Case 1). The log-likelihood used for estimation is the same as the true Data Generating Process. For the error terms, $\varepsilon_t$, in \eqref{errorMEM}, we rely on the Gamma distribution.
	\item \textit{Distributional misspecification} (Case 2). The errors are generated from a Lognormal distribution, namely $\varepsilon_t\sim\text{Ln}(1,\phi^2)$,\footnote{The specification of the probability density function of the Lognormal distribution, having the characteristic specified in \eqref{errorMEM} is provided in \ref{lognormal}.} while for estimation purposes we rely on the misspecified Gamma log-likelihood.
	\item \textit{Structural change in the dispersion parameter} (Case 3 and 5).\footnote{See the experimental setup in \cite{Mackinnon:1985} to evaluate the finite-sample properties of the covariance matrix estimators of a linear regression model with heteroskedastic errors.} The error term follows a Lognormal distribution (misspecification in distribution) and its the dispersion parameter switches between $\phi^2$ for $t=1,\dots,50$, and $\rho^2\phi^2$ for $t=51,\dots,100$, every 50 observations throughout the sample size $T$. We set the parameter that controls the structural change $\rho^2$ to 4 in Case 3 and 2 in Case 5.
	\item \textit{Time-varying dispersion parameter} (Case 4). The baseline MEM, as described in \eqref{errorMEM}-\eqref{condvar}, assumes a constant dispersion parameter $\phi^2$. Here, we relax this assumption by introducing dynamic behavior into the dispersion parameter process,\footnote{\cite{Barreto:2026} allows the dispersion parameter to depend on past values of the dependent variable.} as detailed in \ref{time-varying},\footnote{In the equation describing the dynamics of $\phi_t^2$, the autoregressive coefficient is equal to 0.4, while the intercept term is set to 0.0375, thus ensuring an unconditional value of the dispersion of 0.0625, while the error term follows a Lognormal distribution with a unit dispersion parameter for identifiability.} while the error term follows a Lognormal distribution.
\end{enumerate}
Table~\ref{first} through Table~\ref{fourth} report the Bias, the RMSE of the estimated parameters and the estimated standard errors for each approach, while Table~\ref{five} through Table~\ref{eight} show the Bias and the RMSE  of the estimated standard errors across each approach. Under correct distributional specification (Case 1), the mean of the standard errors based on the proposed estimators of the covariance matrix converges as $T$ increases. For smaller sample sizes $T$, as found in \cite{Fiorentini:1996}, all covariance matrix estimators are very close to each other, except for the OPG estimator, which remains systematically larger. With respect to the RMSE of the estimated standard errors, we find that the IM and GLM estimators show the best performance,\footnote{The IM and GLM estimators differ uniquely in their estimate of $\phi^2$: the former relies on the Gamma assumption, whereas the latter uses the conditional variance assumption in \eqref{GLM}.} while, as expected, the sandwich estimator exhibits greater variability across these cases.
When only the distributional assumption is misspecified (Case 2), the ML estimator of $\phi^2$ is biased, whereas the mean equation parameter estimates remain systematically unbiased \citep[see, e.g.,][]{Mccullagh:2019}. 
Consequently, the H, OPG, and IM estimators suffer from bias (owing to their dependence on the MLE of $\phi^2$), and, more specifically, we find that they tend to slightly underestimate standard errors. Conversely, the sandwich estimator, in the formulation provided by \cite{Engle:Gallo:2006}, is consistent, as the MLE of $\phi^2$'s cancel out.

Finally, when the dispersion parameter is non-constant (Cases 3, 4, and 5), both the sandwich and FR standard error estimators exhibit negligible bias, whereas the commonly adopted GLM estimator is biased, relying on the GLM assumption which assumes a constant ratio between the conditional variance and the squared conditional mean. Notably, The FR estimator often exhibits lower variability than the sandwich estimator, particularly for the dynamic mean parameters. 

\begin{table}[htbp]
\begin{center}
\begin{minipage}{360pt}
\caption{Simulation Results, $T=500$. 10000 Replications.}\label{first}
\begin{tabular}{cccccccccc}
    \hline
    \multicolumn{10}{c}{Case 1} \\
		\hline
		Coef. & Value & Bias & RMSE &  H & OPG & IM & S & GLM & FR   \\ 
		\hline
$\omega$ & 1.500 & 0.183 & 0.508 & 0.450 & 0.459 & 0.450 & 0.449 & 0.450 & 0.447   \\ 
$\alpha_1$ & 0.400 & -0.003 & 0.042 & 0.042 & 0.043 & 0.042 & 0.042 & 0.042 & 0.042  \\ 
$\beta_1$ & 0.500 & -0.007 & 0.051 & 0.051 & 0.052 & 0.051 & 0.051 & 0.051 & 0.050   \\ 
$\gamma_1$ & 0.050 & 0.000 & 0.020 & 0.020 & 0.020 & 0.020 & 0.020 & 0.020 & 0.020   \\ 
$\phi^2$ & 0.062 & -0.000 &   &   &   &   &   &  &    \\ 
		\hline
		\multicolumn{10}{c}{Case 2} \\
		\hline
		Coef. & Value & Bias & RMSE &  H & OPG & IM & S & GLM & FR   \\ 
		\hline
$\omega$ & 1.500 & 0.180 & 0.501 & 0.447 & 0.448 & 0.447 & 0.454 & 0.456 & 0.453   \\ 
$\alpha_1$ & 0.400 & -0.003 & 0.043 & 0.042 & 0.042 & 0.042 & 0.043 & 0.043 & 0.043  \\ 
$\beta_1$ & 0.500 & -0.007 & 0.052 & 0.051 & 0.051 & 0.051 & 0.052 & 0.052 & 0.051  \\ 
$\gamma_1$ & 0.050 & 0.000 & 0.020 & 0.019 & 0.019 & 0.019 & 0.020 & 0.020 & 0.020  \\ 
$\phi^2$ & 0.062 & -0.003 & &   &  &   &   &  &    \\ \hline
		\multicolumn{10}{c}{Case 3} \\
\hline
	Coef. & Value & Bias & RMSE &  H & OPG & IM & S & GLM & FR   \\ 
\hline

$\omega$ & 1.500 & 0.187 & 0.567 & 0.444 & 0.398 & 0.444 & 0.516 & 0.469 & 0.516   \\ 
$\alpha_1$ & 0.400 & -0.003 & 0.053 & 0.045 & 0.040 & 0.045 & 0.053 & 0.047 & 0.052  \\ 
$\beta_1$ & 0.500 & -0.008 & 0.063 & 0.052 & 0.046 & 0.052 & 0.062 & 0.055 & 0.062   \\ 
$\gamma_1$ & 0.050 & 0.000 & 0.032 & 0.030 & 0.029 & 0.030 & 0.031 & 0.031 & 0.031  \\ 
$\phi^2$ & 0.062 & 0.075 &   &  &  &   &   &  &    \\ 	\hline	\multicolumn{10}{c}{Case 4} \\
\hline
	Coef. & Value & Bias & RMSE &  H & OPG & IM & S & GLM & FR   \\  
\hline

$\omega$ & 1.500 & 0.179 & 0.516 & 0.457 & 0.470 & 0.457 & 0.470 & 0.467 & 0.467   \\ 
$\alpha_1$ & 0.400 & -0.004 & 0.060 & 0.043 & 0.034 & 0.043 & 0.058 & 0.043 & 0.058  \\ 
$\beta_1$ & 0.500 & -0.006 & 0.068 & 0.052 & 0.044 & 0.051 & 0.067 & 0.053 & 0.066  \\ 
$\gamma_1$ & 0.050 & 0.000 & 0.020 & 0.019 & 0.019 & 0.019 & 0.020 & 0.020 & 0.020   \\ 
$\phi^2$ & 0.062 & -0.003 &  &  &  &   &   &  &    \\
\hline	\multicolumn{10}{c}{Case 5} \\
	\hline
	Coef. & Value & Bias & RMSE &  H & OPG & IM & S & GLM & FR   \\  
	\hline
$\omega$ & 1.500 & 0.183 & 0.523 & 0.448 & 0.433 & 0.448 & 0.475 & 0.462 & 0.474   \\ 
$\alpha_1$ & 0.400 & -0.003 & 0.046 & 0.043 & 0.041 & 0.043 & 0.046 & 0.044 & 0.046  \\ 
$\beta_1$ & 0.500 & -0.007 & 0.055 & 0.051 & 0.049 & 0.051 & 0.055 & 0.053 & 0.055   \\ 
$\gamma_1$ & 0.050 & 0.000 & 0.024 & 0.024 & 0.023 & 0.024 & 0.024 & 0.024 & 0.024 \\ 
$\phi^2$ & 0.062 & 0.025 &   &   &   &   &   &  &    \\ 
\hline
\end{tabular}%

\footnotesize
\textbf{Notes:} True values, bias, root mean squared error (RMSE), and average estimated standard errors across six estimators: Hessian (H), outer product of gradients (OPG), information matrix (IM), sandwich (S), GLM, and fully robust (FR). See Section~\ref{sim} for the Data Generating Process (DGP) of each Case.
\end{minipage}
\end{center}
\end{table}


\begin{table}[htbp]
\begin{center}
\begin{minipage}{360pt}
\caption{Simulation Results, $T=1000$. 10000 Replications.}\label{second}
	\begin{tabular}{cccccccccc}
   \hline
    \multicolumn{10}{c}{Case 1} \\
		\hline
	Coef. & Value & Bias & RMSE &  H & OPG & IM & S & GLM & FR   \\   
		\hline
$\omega$ & 1.500 & 0.095 & 0.328 & 0.305 & 0.308 & 0.305 & 0.304 & 0.305 & 0.304  \\ 
$\alpha_1$ & 0.400 & -0.001 & 0.030 & 0.030 & 0.030 & 0.030 & 0.030 & 0.030 & 0.029   \\ 
$\beta_1$ & 0.500 & -0.005 & 0.036 & 0.035 & 0.036 & 0.035 & 0.035 & 0.035 & 0.035  \\ 
$\gamma_1$ & 0.050 & 0.000 & 0.014 & 0.014 & 0.014 & 0.014 & 0.014 & 0.014 & 0.014  \\ 
$\phi^2$ & 0.062 & -0.000 &  &   &   &   &   &  &    \\   
		\hline
		\multicolumn{10}{c}{Case 2} \\
		\hline
Coef. & Value & Bias & RMSE &  H & OPG & IM & S & GLM & FR   \\  	\hline		
$\omega$ & 1.500 & 0.090 & 0.327 & 0.302 & 0.300 & 0.303 & 0.308 & 0.309 & 0.308   \\ 
$\alpha_1$ & 0.400 & -0.001 & 0.030 & 0.030 & 0.029 & 0.030 & 0.030 & 0.030 & 0.030  \\ 
$\beta_1$ & 0.500 & -0.004 & 0.036 & 0.035 & 0.035 & 0.035 & 0.036 & 0.036 & 0.036  \\ 
$\gamma_1$ & 0.050 & 0.000 & 0.014 & 0.014 & 0.014 & 0.014 & 0.014 & 0.014 & 0.014   \\ 
$\phi^2$ & 0.062 & -0.003 &   &   &   &   &   &  &    \\ \hline	
\multicolumn{10}{c}{Case 3} \\
\hline
Coef. & Value & Bias & RMSE &  H & OPG & IM & S & GLM & FR   \\  
\hline
$\omega$ & 1.500 & 0.096 & 0.373 & 0.299 & 0.262 & 0.299 & 0.350 & 0.317 & 0.350   \\ 
$\alpha_1$ & 0.400 & -0.001 & 0.038 & 0.032 & 0.028 & 0.032 & 0.037 & 0.034 & 0.037 \\ 
$\beta_1$ & 0.500 & -0.005 & 0.044 & 0.036 & 0.031 & 0.036 & 0.043 & 0.039 & 0.043   \\ 
$\gamma_1$ & 0.050 & 0.000 & 0.022 & 0.021 & 0.020 & 0.021 & 0.022 & 0.022 & 0.022  \\ 
$\phi^2$ & 0.062 & 0.076 &   &   &   &   &   &  &     \\ \hline		\multicolumn{10}{c}{Case 4} \\
\hline
Coef. & Value & Bias & RMSE &  H & OPG & IM & S & GLM & FR   \\  
\hline
$\omega$ & 1.500 & 0.088 & 0.336 & 0.309 & 0.309 & 0.309 & 0.319 & 0.317 & 0.318  \\ 
$\alpha_1$ & 0.400 & -0.002 & 0.044 & 0.030 & 0.023 & 0.030 & 0.042 & 0.031 & 0.042   \\ 
$\beta_1$ & 0.500 & -0.003 & 0.048 & 0.036 & 0.029 & 0.036 & 0.047 & 0.037 & 0.047   \\ 
$\gamma_1$ & 0.050 & 0.000 & 0.014 & 0.013 & 0.013 & 0.013 & 0.014 & 0.014 & 0.014   \\ 
$\phi^2$ & 0.062 & -0.004 &   &   &   &   &   &  &    \\ \hline
\multicolumn{10}{c}{Case 5} \\
\hline
Coef. & Value & Bias & RMSE &  H & OPG & IM & S & GLM & FR   \\  
\hline
$\omega$ & 1.500 & 0.093 & 0.342 & 0.303 & 0.289 & 0.303 & 0.322 & 0.312  & 0.322  \\ 
$\alpha_1$ & 0.400 & -0.001 & 0.033 & 0.030& 0.029 & 0.030  & 0.033& 0.031     & 0.032 \\ 
$\beta_1$ & 0.500  & -0.004 & 0.039 & 0.036 & 0.034 & 0.036& 0.038 & 0.037  & 0.038  \\ 
$\gamma_1$ & 0.050 & 0.000 & 0.017 & 0.017 & 0.016 & 0.017 & 0.017 & 0.017  & 0.017  \\ 
$\phi^2$ & 0.062 & 0.025 &  &  &  &  &  &  &    \\
		\hline
	\end{tabular}

\footnotesize
\textbf{Notes:} True values, bias, root mean squared error (RMSE), and average estimated standard errors across six estimators: Hessian (H), outer product of gradients (OPG), information matrix (IM), sandwich (S), GLM, and fully robust (FR). See Section~\ref{sim} for the Data Generating Process (DGP) of each Case.
\end{minipage}
\end{center}
\end{table}

\begin{table}[htbp]
\begin{center}
\begin{minipage}{360pt}
\caption{Simulation Results, $T=2500$. 10000 Replications.}\label{third}
	\begin{tabular}{cccccccccc}
   \hline
    \multicolumn{10}{c}{Case 1} \\
		\hline
		Coef. & Value & Bias & RMSE &  H & OPG & IM & S & GLM & FR \\ 
		\hline
		
$\omega$ & 1.500 & 0.034 & 0.191 & 0.187 & 0.188 & 0.187 & 0.187 & 0.187 & 0.187  \\ 
$\alpha_1$ & 0.400 & -0.001 & 0.019 & 0.019 & 0.019 & 0.019 & 0.019 & 0.019 & 0.019 \\ 
$\beta_1$ & 0.500 & -0.001 & 0.022 & 0.022 & 0.022 & 0.022 & 0.022 & 0.022 & 0.022 \\ 
$\gamma_1$ & 0.050 & 0.000 & 0.009 & 0.009 & 0.009 & 0.009 & 0.009 & 0.009 & 0.009  \\ 
$\phi^2$ & 0.062 & -0.000 &   &  &  &  & &  &   \\   
		\hline
		\multicolumn{10}{c}{Case 2} \\
		\hline
		Coef. & Value & Bias & RMSE &  H & OPG & IM & S & GLM & FR  \\ 
		\hline
$\omega$ & 1.500 & 0.038 & 0.199 & 0.186 & 0.184 & 0.186 & 0.190 & 0.190 & 0.190  \\ 
$\alpha_1$ & 0.400 & -0.000 & 0.019 & 0.019 & 0.018 & 0.019 & 0.019 & 0.019 & 0.019  \\ 
$\beta_1$ & 0.500 & -0.002 & 0.023 & 0.022 & 0.022 & 0.022 & 0.023 & 0.023 & 0.023  \\ 
$\gamma_1$ & 0.050 & -0.000 & 0.009 & 0.009 & 0.008 & 0.009 & 0.009 & 0.009 & 0.009  \\ 
$\phi^2$ & 0.062 & -0.003 & &  &  &  &  &  &    \\ \hline
		\multicolumn{10}{c}{Case 3} \\
\hline
Coef. & Value & Bias & RMSE &  H & OPG & IM & S & GLM & FR \\ 
\hline
$\omega$ & 1.500 & 0.037 & 0.224 & 0.184 & 0.158 & 0.184 & 0.216 & 0.195 & 0.216  \\ 
$\alpha_1$ & 0.400 & -0.000 & 0.024 & 0.020 & 0.017 & 0.020 & 0.024 & 0.021 & 0.024  \\ 
$\beta_1$ & 0.500 & -0.002 & 0.027 & 0.023 & 0.019 & 0.023 & 0.027 & 0.024 & 0.027  \\ 
$\gamma_1$ & 0.050 & -0.000 & 0.014 & 0.013 & 0.013 & 0.013 & 0.014 & 0.014 & 0.014  \\ 
$\phi^2$ & 0.062 & 0.076 & &  &  &  &  &  &     \\ \hline
\multicolumn{10}{c}{Case 4} \\
\hline
Coef. & Value & Bias & RMSE &  H & OPG & IM & S & GLM & FR \\ 
\hline
$\omega$ & 1.500 & 0.037 & 0.204 & 0.191 & 0.188 & 0.191 & 0.197 & 0.196 & 0.197  \\ 
$\alpha_1$ & 0.400 & -0.001 & 0.028 & 0.019 & 0.014 & 0.019 & 0.027 & 0.019 & 0.027  \\ 
$\beta_1$ & 0.500 & -0.001 & 0.031 & 0.022 & 0.018 & 0.022 & 0.031 & 0.023 & 0.031  \\ 
$\gamma_1$ & 0.050 & -0.000 & 0.009 & 0.008 & 0.008 & 0.008 & 0.009 & 0.009 & 0.009  \\ 
$\phi^2$ & 0.062 & -0.004 &  &  &  &  &  &  &  \\ \hline \multicolumn{10}{c}{Case 5} \\
\hline
		Coef. & Value & Bias & RMSE &  H & OPG & IM & S & GLM & FR  \\ 
\hline
$\omega$ & 1.500 & 0.038 & 0.207 & 0.186 & 0.176 & 0.186 & 0.198 & 0.192 & 0.198  \\
$\alpha_1$ & 0.400 & -0.000 & 0.021 & 0.019 & 0.018 & 0.019 & 0.021 & 0.020 & 0.021 \\ 
$\beta_1$ & 0.500 & -0.002 & 0.024 & 0.022 & 0.021 & 0.022 & 0.024 & 0.023 & 0.024  \\ 
$\gamma_1$ & 0.050 & -0.000 & 0.011 & 0.010 & 0.010 & 0.010 & 0.011 & 0.011 & 0.011  \\ 
$\phi^2$ & 0.062 & 0.025 & & &  &  & &  &   \\ 
		\hline
	\end{tabular}

\footnotesize
\textbf{Notes:} True values, bias, root mean squared error (RMSE), and average estimated standard errors across six estimators: Hessian (H), outer product of gradients (OPG), information matrix (IM), sandwich (S), GLM, and fully robust (FR). See Section~\ref{sim} for the Data Generating Process (DGP) of each Case.
\end{minipage}
\end{center}
\end{table}

\begin{table}[htbp]
\begin{center}
\begin{minipage}{360pt}
\caption{Simulation Results, $T=5000$. 10000 Replications.}\label{fourth}
	\begin{tabular}{cccccccccc}
		\hline
        \multicolumn{10}{c}{Case 1} \\
\hline
		Coef. & Value & Bias & RMSE &  H & OPG & IM & S & GLM & FR  \\ 
		\hline		
$\omega$ & 1.500 & 0.017 & 0.134 & 0.131 & 0.132 & 0.131 & 0.131 & 0.131 & 0.131 \\ 
$\alpha_1$ & 0.400 & -0.000 & 0.013 & 0.013 & 0.013 & 0.013 & 0.013 & 0.013 & 0.013 \\ 
$\beta_1$ & 0.500 & -0.001 & 0.016 & 0.016 & 0.016 & 0.016 & 0.016 & 0.016 & 0.016 \\ 
$\gamma_1$ & 0.050 & 0.000 & 0.006 & 0.006 & 0.006 & 0.006 & 0.006 & 0.006 & 0.006  \\ 
$\phi^2$ & 0.062 & -0.000 & &  &  &  &  &  &    \\ 	\hline
		\multicolumn{10}{c}{Case 2} \\
		\hline
		Coef. & Value & Bias & RMSE &  H & OPG & IM & S & GLM & FR  \\ 
		\hline
		
$\omega$ & 1.500 & 0.018 & 0.137 & 0.131 & 0.128 & 0.131 & 0.133 & 0.133 & 0.133  \\ 
$\alpha_1$ & 0.400 & -0.000 & 0.013 & 0.013 & 0.013 & 0.013 & 0.013 & 0.013 & 0.013  \\ 
$\beta_1$ & 0.500 & -0.001 & 0.016 & 0.016 & 0.015 & 0.016 & 0.016 & 0.016 & 0.016  \\ 
$\gamma_1$ & 0.050 & 0.000 & 0.006 & 0.006 & 0.006 & 0.006 & 0.006 & 0.006 & 0.006  \\ 
$\phi^2$ & 0.062 & -0.003 &  &  &  &  &  &  &    \\ 
\hline
\multicolumn{10}{c}{Case 3} \\
\hline
		Coef. & Value & Bias & RMSE &  H & OPG & IM & S & GLM & FR  \\ 
\hline
$\omega$ & 1.500 & 0.019 & 0.155 & 0.128 & 0.110 & 0.128 & 0.151 & 0.136 & 0.151  \\ 
$\alpha_1$ & 0.400 & -0.000 & 0.017 & 0.014 & 0.012 & 0.014 & 0.017 & 0.015 & 0.017  \\ 
$\beta_1$ & 0.500 & -0.001 & 0.019 & 0.016 & 0.013 & 0.016 & 0.019 & 0.017 & 0.019  \\ 
$\gamma_1$ & 0.050 & 0.000 & 0.010 & 0.009 & 0.009 & 0.009 & 0.010 & 0.010 & 0.010  \\ 
$\phi^2$ & 0.062 & 0.076 &  & &  &  &   &  &  \\ \hline
\multicolumn{10}{c}{Case 4} \\
\hline
		Coef. & Value & Bias & RMSE &  H & OPG & IM & S & GLM & FR  \\ 
\hline
$\omega$ & 1.500 & 0.017 & 0.141 & 0.133 & 0.131 & 0.133 & 0.138 & 0.137 & 0.138  \\ 
$\alpha_1$ & 0.400 & -0.000 & 0.021 & 0.013 & 0.010 & 0.013 & 0.020 & 0.014 & 0.020  \\ 
$\beta_1$ & 0.500 & -0.000 & 0.023 & 0.016 & 0.012 & 0.016 & 0.022 & 0.016 & 0.022  \\ 
$\gamma_1$ & 0.050 & 0.000 & 0.006 & 0.006 & 0.006 & 0.006 & 0.006 & 0.006 & 0.006  \\ 
$\phi^2$ & 0.062 & -0.004 &  &  &  &  &  &  &    \\ 
		\hline
		\multicolumn{10}{c}{Case 5} \\
		\hline
		Coef. & Value & Bias & RMSE &  H & OPG & IM & S & GLM & FR  \\ 
		\hline
$\omega$ & 1.500 & 0.018 & 0.143 & 0.130 & 0.122 & 0.130 & 0.139 & 0.135 & 0.139 \\ 
$\alpha_1$ & 0.400 & -0.000 & 0.014 & 0.014 & 0.013 & 0.014 & 0.015 & 0.014 & 0.015  \\ 
$\beta_1$ & 0.500 & -0.001 & 0.017 & 0.016 & 0.015 & 0.016 & 0.017 & 0.016 & 0.017  \\ 
$\gamma_1$ & 0.050 & 0.000 & 0.008 & 0.007 & 0.007 & 0.007 & 0.008 & 0.008 & 0.008  \\ 
$\phi^2$ & 0.062 & 0.025 & &  & &  &    &  &  \\ \hline
	\end{tabular}

\footnotesize
\textbf{Notes:} True values, bias, root mean squared error (RMSE), and average estimated standard errors across six estimators: Hessian (H), outer product of gradients (OPG), information matrix (IM), sandwich (S), GLM, and fully robust (FR). See Section~\ref{sim} for the Data Generating Process (DGP) of each Case.
\end{minipage}
\end{center}
\end{table}

\begin{table}[htbp]
	\caption{Simulation Results: estimated Standard Errors, $T=500$. 10000 Replications.}\label{five}
	\resizebox{\textwidth}{!}{\begin{tabular}{cc|cccccc|cccccc}
			\hline
			& & \multicolumn{6}{c|}{Bias} &
			\multicolumn{6}{c}{RMSE} \\
            		\hline
			\multicolumn{14}{c}{Case 1} \\
			\hline
			Coef.& Value &  H & OPG & IM & S & GLM & FR & H & OPG & IM & S & GLM & FR \\ 
			\hline
$\omega$ & 1.500 & -2.385 & -1.541 & -2.382 & -2.536 & -2.390 & -2.696 & 9.491 & 9.492 & 9.253 & 10.286 & 9.257 & 9.554  \\ 
$\alpha_1$ & 0.400 & -0.041 & 0.030 & -0.045 & -0.041 & -0.045 & -0.071 & 0.274 & 0.246 & 0.154 & 0.520 & 0.161 & 0.260  \\ 
$\beta_1$ & 0.500 & -0.012 & 0.084 & -0.007 & -0.015 & -0.007 & -0.038 & 0.555 & 0.502 & 0.414 & 0.883 & 0.417 & 0.481 \\ 
$\gamma_1$ & 0.050 & 0.001 & 0.022 & -0.004 & 0.003 & -0.005 & -0.010 & 0.107 & 0.127 & 0.105 & 0.125 & 0.106 & 0.122  \\ 
		\hline
			\multicolumn{14}{c}{Case 2} \\
			\hline
	Coef.& Value &  H & OPG & IM & S & GLM & FR & H & OPG & IM & S & GLM & FR\\ 
			\hline
$\omega$ & 1.500 & -2.088 & -1.998 & -2.059 & -1.420 & -1.184 & -1.506 &  9.150 & 9.314 & 9.018 & 9.802 & 9.035 & 9.346  \\ 
$\alpha_1$ & 0.400 & -0.104 & -0.106 & -0.108 & -0.021 & -0.025 & -0.051 &  0.300 & 0.279 & 0.193 & 0.551 & 0.176 & 0.291   \\ 
$\beta_1$ & 0.500 & -0.063 & -0.052 & -0.054 & 0.029 & 0.045 & 0.014 &  0.546 & 0.499 & 0.415 & 0.886 & 0.426 & 0.495 \\ 
$\gamma_1$ & 0.050 & -0.045 & -0.058 & -0.050 & -0.004 & -0.012 & -0.018 &  0.116 & 0.134 & 0.116 & 0.131 & 0.112 & 0.132   \\ 
		\hline
			\multicolumn{14}{c}{Case 3} \\
			\hline
		Coef.& Value &  H & OPG & IM & S & GLM & FR & H & OPG & IM & S & GLM & FR  \\ 
			\hline
$\omega$ & 1.500 & -9.116 & -13.719 & -9.113 & -1.902 & -6.625 & -1.976 & 13.866 & 16.835 & 13.639 & 13.889 & 12.720 & 13.214  \\ 
$\alpha_1$ & 0.400 & -0.800 & -1.278 & -0.814 & -0.043 & -0.562 & -0.095 & 0.891 & 1.333 & 0.850 & 0.915 & 0.646 & 0.621  \\ 
$\beta_1$ & 0.500 & -0.965 & -1.576 & -0.965 & -0.023 & -0.673 & -0.057 &1.208 & 1.703 & 1.099 & 1.410 & 0.884 & 0.841 \\ 
$\gamma_1$ & 0.050 & -0.182 & -0.244 & -0.193 & -0.022 & -0.027 & -0.046 & 0.269 & 0.337 & 0.272 & 0.301 & 0.249 & 0.315 \\ 
	\hline
		\multicolumn{14}{c}{Case 4} \\
		\hline
	Coef.& Value &  H & OPG & IM & S & GLM & FR & H & OPG & IM & S & GLM & FR \\ 
		\hline
$\omega$ & 1.500 & -2.719 & -1.499 & -2.743 & -1.471 & -1.736 & -1.768 & 11.139 & 48.950 & 10.599 & 12.469 & 9.812 & 11.297 \\ 
$\alpha_1$ & 0.400 & -1.722 & -2.613 & -1.745 & -0.177 & -1.650 & -0.244 &  1.781 & 2.680 & 1.832 & 1.634 & 1.696 & 1.945 \\ 
$\beta_1$ & 0.500 & -1.579 & -2.325 & -1.592 & -1.479 & -0.139 & -0.073 & 1.752 & 3.891 & 1.748 & 1.832 & 1.572 & 1.841 \\ 
$\gamma_1$ & 0.050 & -0.109 & -0.102 & -0.114 & -0.018 & -0.072 & -0.030 &  0.364 & 2.505 & 0.375 & 0.258 & 0.250 & 0.278 \\ 
	\hline
\multicolumn{14}{c}{Case 5} \\
\hline
	Coef.& Value &  H & OPG & IM & S & GLM & FR & H & OPG & IM & S & GLM & FR \\ 
\hline
$\omega$ & 1.500 & -4.232 & -5.718 & -4.204 & -1.536 & -2.826 & -1.609 &  10.319 & 11.023 & 10.151 & 10.971 & 9.962 & 10.476 \\ 
$\alpha_1$ & 0.400 & -0.316 & -0.488 & -0.321 & -0.020 & -0.189 & -0.053 & 0.446 & 0.575 & 0.372 & 0.657 & 0.285 & 0.395 \\ 
$\beta_1$ & 0.500 & -0.335 & -0.534 & -0.327 & 0.018 & -0.169 & 0.002 &  0.686 & 0.769 & 0.556 & 1.038 & 0.500 & 0.605 \\ 
$\gamma_1$ & 0.050 & -0.083 & -0.112 & -0.090 & -0.007 & -0.018 & -0.025& 0.161 & 0.194 & 0.162 & 0.181 & 0.153 & 0.185  \\ 
			\hline
\end{tabular}}
\footnotesize
\textbf{Notes:} True values, bias and root mean squared error (RMSE) of estimated standard errors across six estimators: Hessian (H), outer product of gradients (OPG), information matrix (IM), sandwich (S), GLM, and fully robust (FR). See Section~\ref{sim} for the Data Generating Process (DGP) of each Case. The values are expressed in percentage terms.
\end{table}

\begin{table}[htbp]
	\caption{Simulation Results: estimated Standard Errors, $T=1000$. 10000 Replications.}\label{six}
\resizebox{\textwidth}{!}{\begin{tabular}{cc|cccccc|cccccc}
			\hline
			& & \multicolumn{6}{c|}{Bias} &
			\multicolumn{6}{c}{RMSE} \\
            	\hline
\multicolumn{14}{c}{Case 1} \\
			\hline
		Coef.& Value &  H & OPG & IM & S & GLM & FR & H & OPG & IM & S & GLM & FR  \\ 
			\hline
$\omega$ & 1.500 & -0.890 & -0.590 & -0.892 & -0.956 & -0.898 & -1.010 &  4.366 & 4.431 & 4.290 & 4.629 & 4.291 & 4.411  \\ 
$\alpha_1$ & 0.400 & -0.038 & -0.017 & -0.041 & -0.034 & -0.041 & -0.047& 0.140 & 0.124 & 0.084 & 0.260 & 0.087 & 0.135\\ 
$\beta_1$ & 0.500 & -0.038 & -0.006 & -0.038 & -0.037 & -0.038 & -0.048 &  0.263 & 0.233 & 0.197 & 0.414 & 0.199 & 0.233 \\ 
$\gamma_1$ & 0.050 & -0.014 & -0.007 & -0.016 & -0.013 & -0.016 & -0.017 &  0.055 & 0.062 & 0.055 & 0.064 & 0.056 & 0.064  \\ 
\hline
			\multicolumn{14}{c}{Case } \\ \hline
		Coef.& Value &  H & OPG & IM & S & GLM & FR & H & OPG & IM & S & GLM & FR  \\ 
			\hline
$\omega$ & 1.500 & -1.219 & -1.472 & -1.207 & -0.699 & -0.599 & -0.715 & 4.381 & 4.505 & 4.324 & 4.626 & 4.285 & 4.462 \\ 
$\alpha_1$ & 0.400 & -0.086 & -0.114 & -0.086 & -0.028 & -0.027 & -0.036 &  0.165 & 0.175 & 0.118 & 0.272 & 0.090 & 0.150  \\ 
$\beta_1$ & 0.500 & -0.075 & -0.103 & -0.071 & -0.011 & 0.000 & -0.012 & 0.272 & 0.259 & 0.209 & 0.427 & 0.202 & 0.244  \\ 
$\gamma_1$ & 0.050 & -0.023 & -0.041 & -0.025 & 0.005 & 0.003 & 0.001 & 0.058 & 0.073 & 0.058 & 0.065 & 0.056 & 0.066  \\  
			\hline			\multicolumn{14}{c}{Case 3} \\ \hline
		Coef.& Value &  H & OPG & IM & S & GLM & FR & H & OPG & IM & S & GLM & FR  \\ 
			\hline
$\omega$ & 1.500 & -6.077 & -9.786 & -6.082 & -1.019 & -4.327 & -1.038 &  7.809 & 10.783 & 7.741 & 6.603 & 6.726 & 6.424 \\ 
$\alpha_1$ & 0.400 & -0.603 & -1.001 & -0.608 & -0.046 & -0.422 & -0.060 &  0.634 & 1.021 & 0.620 & 0.453 & 0.451 & 0.333 \\ 
$\beta_1$ & 0.500 & -0.707 & -1.208 & -0.707 & -0.025 & -0.494 & -0.034 & 0.789 & 1.249 & 0.751 & 0.679 & 0.567 & 0.434 \\ 
$\gamma_1$ & 0.050 & -0.116 & -0.190 & -0.120 & -0.002 & 0.002 & -0.010 &0.152 & 0.223 & 0.154 & 0.150 & 0.123 & 0.158  \\ 
			\hline			\multicolumn{14}{c}{Case 4} \\ \hline
		Coef.& Value &  H & OPG & IM & S & GLM & FR & H & OPG & IM & S & GLM & FR  \\ 
\hline
$\omega$ & 1.500 & -1.580 & -1.604 & -1.578 & -0.619 & -0.783 & -0.710 & 4.727 & 4.820 & 4.600 & 8.060 & 5.020 & 5.776 \\ 
$\alpha_1$ & 0.400 & -1.337 & -2.058 & -1.346 & -0.129 & -1.269 & -0.146& 1.350 & 2.069 & 1.352 & 1.108 & 1.289 & 1.112\\ 
$\beta_1$ & 0.500 & -1.252 & -1.894 & -1.256 & -0.085 & -1.166 & -0.094 &1.288 & 1.912 & 1.273 & 1.053 & 1.194 & 1.027 \\ 
$\gamma_1$ & 0.050 & -0.081 & -0.114 & -0.083 & -0.008 & -0.048 & -0.012  & 0.117 & 0.143 & 0.117 & 0.267 & 0.145 & 0.165\\ 
			\hline			\multicolumn{14}{c}{Case 5} \\ \hline
		Coef.& Value &  H & OPG & IM & S & GLM & FR & H & OPG & IM & S & GLM & FR  \\ 
\hline
$\omega$ & 1.500 & -2.681 & -4.097 & -2.672 & -0.767 & -1.712 & -0.785 &5.184 & 6.005 & 5.115 & 5.216 & 4.834 & 5.038  \\ 
$\alpha_1$ & 0.400 & -0.247 & -0.409 & -0.248 & -0.031 & -0.152 & -0.040 & 0.293 & 0.438 & 0.266 & 0.325 & 0.186 & 0.207 \\ 
$\beta_1$ & 0.500 & -0.262 & -0.451 & -0.259 & -0.008 & -0.146 & -0.011 & 0.391 & 0.525 & 0.336 & 0.502 & 0.269 & 0.303  \\ 
$\gamma_1$ & 0.050 & -0.048 & -0.083 & -0.051 & 0.004 & 0.002 & -0.002 &  0.084 & 0.114 & 0.084 & 0.090 & 0.075 & 0.092 \\
			\hline
	\end{tabular}}
\footnotesize
\textbf{Notes:} True values, bias and root mean squared error (RMSE) of estimated standard errors across six estimators: Hessian (H), outer product of gradients (OPG), information matrix (IM), sandwich (S), GLM, and fully robust (FR). See Section~\ref{sim} for the Data Generating Process (DGP) of each Case. The values are expressed in percentage terms.   
\end{table}

\begin{table}[htbp]
	\caption{Simulation Results: estimated Standard Errors, $T=2500$. 10000 Replications.}\label{seven}
	\resizebox{\textwidth}{!}{\begin{tabular}{cc|cccccc|cccccc}
			\hline
			& & \multicolumn{6}{c|}{Bias} &
			\multicolumn{6}{c}{RMSE} \\
\hline
\multicolumn{14}{c}{Case 1} \\ \hline
	Coef.& Value &  H & OPG & IM & S & GLM & FR & H & OPG & IM & S & GLM & FR \\ 
			\hline
$\omega$ & 1.500 & -0.047 & 0.039 & -0.043 & -0.076 & -0.043 & -0.079  & 1.614 & 1.657 & 1.593 & 1.708 & 1.595 & 1.642  \\ 
$\alpha_1$ & 0.400 & -0.006 & 0.000 & -0.007 & -0.006 & -0.007 & -0.009 &  0.054 & 0.050 & 0.030 & 0.100 & 0.031 & 0.051 \\ 
$\beta_1$ & 0.500 & -0.003 & 0.007 & -0.002 & -0.004 & -0.002 & -0.005  & 0.099 & 0.089 & 0.074 & 0.158 & 0.074 & 0.088 \\ 
$\gamma_1$ & 0.050 & -0.008 & -0.006 & -0.009 & -0.008 & -0.009 & -0.009 &  0.023 & 0.026 & 0.023 & 0.026 & 0.023 & 0.026  \\ 
\hline
\multicolumn{14}{c}{Case 2} \\ \hline
Coef.& Value &  H & OPG & IM & S & GLM & FR & H & OPG & IM & S & GLM & FR \\ 
			\hline
$\omega$ & 1.500 & -0.908 & -1.185 & -0.902 & -0.558 & -0.525 & -0.557  & 1.884 & 2.047 & 1.864 & 1.868 & 1.746 & 1.816  \\ 
$\alpha_1$ & 0.400 & -0.056 & -0.084 & -0.055 & -0.020 & -0.018 & -0.020 &  0.079 & 0.100 & 0.064 & 0.110 & 0.039 & 0.062  \\ 
$\beta_1$ & 0.500 & -0.056 & -0.088 & -0.054 & -0.014 & -0.009 & -0.013 &  0.117 & 0.129 & 0.094 & 0.169 & 0.079 & 0.096 \\ 
$\gamma_1$ & 0.050 & -0.021 & -0.035 & -0.021 & -0.003 & -0.004 & -0.004 & 0.030 & 0.043 & 0.030 & 0.026 & 0.022 & 0.027  \\ 
\hline			\multicolumn{14}{c}{Case 3} \\ \hline
Coef.& Value &  H & OPG & IM & S & GLM & FR & H & OPG & IM & S & GLM & FR \\ 
\hline
$\omega$ & 1.500 & -3.756 & -6.318 & -3.757 & -0.532 & -2.652 & -0.533  & 4.209 & 6.561 & 4.195 & 2.578 & 3.329 & 2.561  \\ 
$\alpha_1$ & 0.400 & -0.386 & -0.665 & -0.387 & -0.023 & -0.266 & -0.025 & 0.394 & 0.670 & 0.390 & 0.186 & 0.274 & 0.138  \\ 
$\beta_1$ & 0.500 & -0.463 & -0.812 & -0.462 & -0.020 & -0.325 & -0.020 & 0.482 & 0.823 & 0.472 & 0.268 & 0.343 & 0.176 \\ 
$\gamma_1$ & 0.050 & -0.089 & -0.150 & -0.090 & -0.014 & -0.010 & -0.016 &  0.097 & 0.157 & 0.098 & 0.063 & 0.052 & 0.067 \\ 
\hline		\multicolumn{14}{c}{Case 4} \\ \hline
Coef.& Value &  H & OPG & IM & S & GLM & FR & H & OPG & IM & S & GLM & FR \\ 
\hline
$\omega$ & 1.500 & -1.050 & -1.283 & -1.048 & -0.446 & -0.530 & -0.456  & 2.031 & 2.210 & 1.989 & 2.324 & 1.997 & 2.224 \\ 
$\alpha_1$ & 0.400 & -0.926 & -1.435 & -0.929 & -0.089 & -0.877 & -0.088 &  0.930 & 1.440 & 0.931 & 0.603 & 0.883 & 0.679  \\ 
$\beta_1$ & 0.500 & -0.874 & -1.333 & -0.875 & -0.063 & -0.814 & -0.061  & 0.882 & 1.339 & 0.879 & 0.588 & 0.823 & 0.623  \\ 
$\gamma_1$ & 0.050 & -0.056 & -0.088 & -0.057 & -0.008 & -0.033 & -0.007 &  0.066 & 0.094 & 0.066 & 0.075 & 0.064 & 0.086  \\ 
\hline \multicolumn{14}{c}{Case 5} \\ \hline
Coef.& Value &  H & OPG & IM & S & GLM & FR & H & OPG &IM & S & GLM & FR \\ 
\hline
$\omega$ & 1.500 & -1.759 & -2.811 & -1.756 & -0.528 & -1.156 & -0.529  & 2.469 & 3.297 & 2.451 & 2.068 & 2.116 & 2.026  \\ 
$\alpha_1$ & 0.400 & -0.161 & -0.279 & -0.161 & -0.021 & -0.099 & -0.022  & 0.173 & 0.287 & 0.165 & 0.133 & 0.108 & 0.086  \\ 
$\beta_1$ & 0.500 & -0.181 & -0.322 & -0.180 & -0.017 & -0.108 & -0.015  & 0.214 & 0.339 & 0.198 & 0.198 & 0.139 & 0.121  \\ 
$\gamma_1$ & 0.050 & -0.040 & -0.068 & -0.041 & -0.007 & -0.007 & -0.008 &  0.048 & 0.075 & 0.049 & 0.037 & 0.031 & 0.038  \\ 
\hline
	\end{tabular}}
\footnotesize
\textbf{Notes:} True values, bias and root mean squared error (RMSE) of estimated standard errors across six estimators: Hessian (H), outer product of gradients (OPG), information matrix (IM), sandwich (S), GLM, and fully robust (FR). See Section~\ref{sim} for the Data Generating Process (DGP) of each Case. The values are expressed in percentage terms. 
\end{table}

\begin{table}[htbp]
	\caption{Simulation Results: estimated Standard Errors, $T=5000$. 10000 Replications.}\label{eight}
	\resizebox{\textwidth}{!}{\begin{tabular}{cc|cccccc|cccccc}
			\hline
			& & \multicolumn{6}{c|}{Bias} &
			\multicolumn{6}{c}{RMSE} \\
	\hline		\multicolumn{14}{c}{Case 1} \\ \hline
	Coef.& Value &  H & OPG & IM & S & GLM & FR & H & OPG & IM & S & GLM & FR \\ 
\hline
$\omega$ & 1.500 & -0.200 & -0.174 & -0.198 & -0.205 & -0.198 & -0.207 &  0.836 & 0.852 & 0.825 & 0.877 & 0.826 & 0.850  \\ 
$\alpha_1$ & 0.400 & 0.003 & 0.005 & 0.003 & 0.003 & 0.003 & 0.002  & 0.027 & 0.025 & 0.015 & 0.050 & 0.016 & 0.026  \\ 
$\beta_1$ & 0.500 & 0.003 & 0.006 & 0.003 & 0.002 & 0.003 & 0.002 &  0.050 & 0.045 & 0.037 & 0.079 & 0.038 & 0.045 \\ 
$\gamma_1$ & 0.050 & -0.004 & -0.004 & -0.004 & -0.004 & -0.004 & -0.005 &  0.012 & 0.013 & 0.011 & 0.013 & 0.012 & 0.013  \\ 
\hline		\multicolumn{14}{c}{Case 2} \\ \hline
Coef.& Value &  H & OPG & IM & S & GLM & FR & H & OPG & IM & S & GLM & FR \\ 
		\hline
$\omega$ & 1.500 & -0.482 & -0.712 & -0.482 & -0.223 & -0.217 & -0.228  & 0.944 & 1.086 & 0.935 & 0.904 & 0.846 & 0.879  \\ 
$\alpha_1$ & 0.400 & -0.025 & -0.048 & -0.025 & 0.002 & 0.002 & 0.001 & 0.037 & 0.055 & 0.030 & 0.054 & 0.017 & 0.030 \\ 
$\beta_1$ & 0.500 & -0.030 & -0.058 & -0.030 & 0.002 & 0.002 & 0.001 & 0.059 & 0.074 & 0.048 & 0.083 & 0.039 & 0.047  \\ 
$\gamma_1$ & 0.050 & -0.010 & -0.022 & -0.010 & 0.002 & 0.002 & 0.002 & 0.015 & 0.025 & 0.015 & 0.013 & 0.011 & 0.013  \\ 
\hline		\multicolumn{14}{c}{Case 3} \\ \hline
Coef.& Value &  H & OPG & IM & S & GLM & FR & H & OPG & IM & S & GLM & FR \\ 
\hline
$\omega$ & 1.500 & -2.525 & -4.394 & -2.525 & -0.236 & -1.742 & -0.236& 2.693 & 4.480 & 2.686 & 1.267 & 2.003 & 1.253  \\ 
$\alpha_1$ & 0.400 & -0.255 & -0.461 & -0.256 & 0.006 & -0.169 & 0.005 & 0.258 & 0.463 & 0.257 & 0.093 & 0.173 & 0.070  \\ 
$\beta_1$ & 0.500 & -0.319 & -0.576 & -0.319 & -0.001 & -0.221 & -0.002 & 0.326 & 0.580 & 0.323 & 0.135 & 0.228 & 0.090  \\ 
$\gamma_1$ & 0.050 & -0.049 & -0.096 & -0.049 & 0.006 & 0.008 & 0.005 & 0.053 & 0.099 & 0.053 & 0.032 & 0.027 & 0.033  \\ 
\hline		\multicolumn{14}{c}{Case 4} \\ \hline
Coef.& Value &  H & OPG & IM & S & GLM & FR & H & OPG & IM & S & GLM & FR \\ 
\hline
$\omega$ & 1.500 & -0.682 & -0.914 & -0.683 & -0.236 & -0.295 & -0.244 &  1.102 & 1.290 & 1.083 & 1.184 & 1.102 & 1.136  \\ 
$\alpha_1$ & 0.400 & -0.716 & -1.097 & -0.717 & -0.075 & -0.678 & -0.070 & 0.717 & 1.100 & 0.717 & 0.479 & 0.681 & 0.568  \\ 
$\beta_1$ & 0.500 & -0.677 & -1.022 & -0.678 & -0.057 & -0.632 & -0.053 &  0.680 & 1.025 & 0.679 & 0.455 & 0.637 & 0.523 \\ 
$\gamma_1$ & 0.050 & -0.036 & -0.062 & -0.036 & 0.002 & -0.018 & 0.002 & 0.040 & 0.064 & 0.040 & 0.049 & 0.039 & 0.057  \\  
			\hline \multicolumn{14}{c}{Case 5} \\ \hline
	Coef.& Value &  H & OPG & IM & S & GLM & FR & H & OPG & IM & S & GLM & FR \\  
\hline
$\omega$ & 1.500 & -1.120 & -1.906 & -1.120 & -0.241 & -0.697 & -0.243 & 1.409 & 2.087 & 1.401 & 1.014 & 1.116 & 0.991\\ 
			$\alpha_1$ & 0.400 & -0.097 & -0.186 & -0.097 & 0.005 & -0.053 & 0.004 &  0.101 & 0.189 & 0.099 & 0.065 & 0.057 & 0.042 \\ 
			$\beta_1$ & 0.500 & -0.118 & -0.224 & -0.118 & 0.001 & -0.067 & -0.000 & 0.131 & 0.231 & 0.125 & 0.099 & 0.080 & 0.060  \\ 
			$\gamma_1$ & 0.050 & -0.020 & -0.042 & -0.020 & 0.004 & 0.004 & 0.003 & 0.024 & 0.045 & 0.024 & 0.019 & 0.016 & 0.019  \\ 
            \hline
\end{tabular}}
\footnotesize
\textbf{Notes:} True values, bias and root mean squared error (RMSE) of estimated standard errors across six estimators: Hessian (H), outer product of gradients (OPG), information matrix (IM), sandwich (S), GLM, and fully robust (FR). See Section~\ref{sim} for the Data Generating Process (DGP) of each Case. The values are expressed in percentage terms.   
\end{table}

\newpage
\newpage
\section{Empirical Analysis}\label{sec:empr}
In the real data application, we consider the realized volatility series of the 30 constituent stocks of DJIA, namely: Apple Inc. (AAPL), Amgen Inc. (AMGN), Amazon.com Inc. (AMZN), American Express Company (AXP), The Boeing Company (BA), Caterpillar Inc. (CAT), Salesforce Inc. (CRM), Cisco Systems Inc. (CSCO), Chevron Corporation (CVX), The
Walt Disney Company (DIS), Dow Inc. (DOW), The Goldman Sachs Group Inc. (GS), The Home Depot Inc. (HD), Honeywell International Inc. (HON), International Business Machines
(IBM), Intel Corporation (INTC), Johnson \& Johnson (JNJ), JPMorgan Chase \& Co. (JPM), The Coca-Cola Company (KO), McDonald’s Corporation (MCD), 3M Company (MMM), Merck \& Co. Inc. (MRK), Microsoft Corporation (MSFT), NIKE Inc.
(NKE), The Procter \& Gamble Company (PG), The Travelers Companies Inc. (TRV), United Health Group Inc. (UNH), Visa Inc. (V), Verizon Communications Inc. (VZ), Walmart Inc. (WMT). Daily data for each stock were retrieved from the VOLARE platform \citep{Cipollini:2026}, covering the period
from January 2, 2015, to January 30, 2026. For our purposes, we consider the percent annualized realized volatility, defined as $\sqrt{\text{RV}_t}\times\sqrt{252}\times100$, where $\text{RV}_t$ is the plain vanilla realized variance at date $t$, computed from 5-minute intraday returns \citep{Andersen:Bollerslev:1998}. For each volatility series, we estimate both an AMEM(1,1) and an AMEM(2,1), whose results are summarized in Tables~\ref{summary_est_mem}-\ref{summary_est_mem21}. 

For the MEM(1,1), the sign and the magnitude of the estimated parameters align with expectations: the average estimated value of $\beta_1$ is slightly larger than the average $\alpha_1$, while the estimated $\gamma_1$'s, capturing the asymmetric effect, are generally close to 0. The estimated volatility persistence, computed as $\alpha_1 + \beta_1 + \gamma_1/2$, is high across all series, ranging from $0.862$ to $0.956$. 

The AMEM(2,1) parameterization yields a negative average estimate for $\alpha_2$, capturing the effect of the second lag of the dependent variable. This allows the model to accommodate a more rapid decay of past volatility bursts compared to the AMEM(1,1).\footnote{See also the estimation results in \cite{Cipollini:2026}.} Additionally, the AMEM(2,1) exhibits higher average values for both $\beta_1$ and overall volatility persistence. 

For both specifications, the average estimate of $\phi^2$ obtained via the GLM variance-based moment condition exceeds the corresponding MLE estimate, provides evidence suggestive of misspecification of the conditional Gamma distribution. Nevertheless, according to the BIC reported in Table~\ref{BIC}, the AMEM(2,1) is the preferred parameterization for all volatility series.

Focusing on standard errors for the selected specification, namely the AMEM(2,1),\footnote{Standard errors relative to the AMEM(1,1) are available upon request.} the Hessian estimates lie between the OPG and Sandwich estimates \citep[see, e.g.,][]{Boldea:2009}. Comparing covariance matrix estimators robust to distributional misspecification, namely the Sandwich, GLM, and FR estimators, we observe that the Sandwich estimator yields the largest standard errors. While the GLM and FR standard errors for the dynamic coefficients are very close to each other, the FR approach yields smaller standard errors for the intercept term in 25 out of 30 cases. Based on these empirical results, we suggest reporting both GLM and FR standard errors.

\begin{table}[htbp]
\begin{center}
\begin{minipage}{255pt}
\caption{Annualized realized volatility, January 2015 - January 2026. Parameter estimates of the AMEM(1,1) models for the 30 constituents of the Dow Jones Industrial Average index.}\label{summary_est_mem}
\begin{tabular}{ccccccc}
  \hline
 Coef. & Min. & Q1 & Q2 & Mean & Q3 & Max. \\ 
  \hline
$\omega$ & 0.942 & 1.446 & 1.669 & 1.825 & 2.140 & 3.328 \\ 
  $\alpha_1$ & 0.346 & 0.392 & 0.421 & 0.414 & 0.432 & 0.526 \\ 
  $\beta_1$ & 0.303 & 0.451 & 0.480 & 0.481 & 0.513 & 0.595 \\ 
  $\gamma_1$ & -0.007 & 0.012 & 0.024 & 0.024 & 0.032 & 0.082 \\ 
  $\phi^2_{MLE}$ & 0.062 & 0.074 & 0.079 & 0.080 & 0.084 & 0.137 \\ 
  $\phi^2_{MOM}$ & 0.075 & 0.097 & 0.103 & 0.114 & 0.118 & 0.267 \\ 
  Pers. & 0.862 & 0.892 & 0.905 & 0.907 & 0.923 & 0.956 \\ 
   \hline
 \end{tabular}
 
\footnotesize {\textbf{Notes:}  $\phi^2_{MLE}$ and $\phi^2_{MOM}$ are the estimates of $\phi^2$ through ML and the method of moments, respectively, while Pers. is the estimated volatility persistence.  The summary statistics are the minimum (Min.), the first quartile (Q1), the median (Q2), the Mean, the third quartile (Q3), the maximum
(Max.) of the parameter estimates.}
\end{minipage}
\end{center}
\end{table}

\begin{table}[htbp]
\begin{center}
\begin{minipage}{270pt}
\caption{Annualized realized volatility, January 2015 - January 2026. Parameter estimates of the AMEM(2,1) models for the 30 constituents of the Dow Jones Industrial Average index.}\label{summary_est_mem21}
\begin{tabular}{ccccccc}
  \hline
Coef. & Min. & Q1 & Q2 & Mean & Q3 & Max. \\ 
  \hline
$\omega$ & 0.178 & 0.352 & 0.535 & 0.507 & 0.660 & 1.142 \\ 
  $\alpha_1$ & 0.389 & 0.442 & 0.467 & 0.463 & 0.482 & 0.556 \\ 
  $\alpha_2$ & -0.426 & -0.343 & -0.318 & -0.305 & -0.261 & -0.161 \\ 
  $\beta_1$ & 0.706 & 0.777 & 0.805 & 0.808 & 0.846 & 0.895 \\ 
  $\gamma_1$ & -0.005 & 0.010 & 0.016 & 0.015 & 0.021 & 0.037 \\ 
  $\phi^2_{MLE}$ & 0.062 & 0.073 & 0.077 & 0.079 & 0.083 & 0.136 \\ 
  $\phi^2_{MOM}$ & 0.075 & 0.095 & 0.102 & 0.113 & 0.116 & 0.274 \\ 
 Pers. &  0.946  & 0.969 & 0.973 & 0.973  & 0.983 & 0.991 \\
   \hline
\end{tabular}
 
\footnotesize {\textbf{Notes:}   $\phi^2_{MLE}$ and $\phi^2_{MOM}$ are the estimates of $\phi^2$ through ML and the method of moments, respectively, while Pers. is the estimated volatility persistence.   The summary statistics are the minimum (Min.), the first quartile (Q1), the median (Q2), the Mean, the third quartile (Q3), the maximum
(Max.) of the parameter estimates.}
\end{minipage}
\end{center}
\end{table}

\begin{table}[htbp]\label{BIC}
\begin{center}
\begin{minipage}{170pt}
\caption{Annualized realized volatility, January 2015 - January 2026. Bayesian Information Criterion (BIC) of the AMEM(2,1) models for the 30 constituents of the Dow Jones Industrial Average index.}\label{BIC}
\centering
\begin{tabular}{ccc}
  \hline
 Symbol & AMEM(1,1) & AMEM(2,1) \\ 
  \hline
AAPL & 6.201 & 6.192 \\ 
  AMGN & 6.339 & 6.330 \\ 
  AMZN & 6.996 & 6.991 \\ 
  AXP & 6.108 & 6.079 \\ 
  BA & 6.712 & 6.679 \\ 
  CAT & 6.316 & 6.307 \\ 
  CRM & 6.710 & 6.702 \\ 
  CSCO & 5.886 & 5.871 \\ 
  CVX & 5.900 & 5.894 \\ 
  DIS & 6.107 & 6.093 \\ 
  GS & 6.163 & 6.154 \\ 
  HD & 5.922 & 5.909 \\ 
  HON & 5.783 & 5.750 \\ 
  IBM & 5.785 & 5.764 \\ 
  JNJ & 5.639 & 5.637 \\ 
  JPM & 5.974 & 5.964 \\ 
  KO & 5.234 & 5.223 \\ 
  MCD & 5.564 & 5.558 \\ 
  MMM & 5.943 & 5.923 \\ 
  MRK & 5.976 & 5.960 \\ 
  MSFT & 6.263 & 6.252 \\ 
  NKE & 6.211 & 6.193 \\ 
  NVDA & 7.166 & 7.157 \\ 
  PG & 5.422 & 5.418 \\ 
  SHW & 6.198 & 6.170 \\ 
  TRV & 5.719 & 5.703 \\ 
  UNH & 6.277 & 6.257 \\ 
  V & 5.974 & 5.968 \\ 
  VZ & 5.670 & 5.663 \\ 
  WMT & 5.755 & 5.747 \\ 
   \hline
\end{tabular}
\end{minipage}
\end{center}
\end{table}

\begin{table}[htbp]
\begin{center}
\begin{minipage}{250pt}
\caption{Annualized realized volatility, January 2015 - January 2026. Estimated Standard Errors of the parameter $\omega$ across six estimators: Hessian (H), outer product of gradients (OPG), information matrix (IM), sandwich (S), GLM, and fully robust (FR).}\label{omega}
\begin{tabular}{ccccccc}
  \hline
Symbol & H & OPG & IM & S & GLM & FR \\ 
  \hline
AAPL & 0.153 & 0.091 & 0.102 & 0.277 & 0.114 & 0.118 \\ 
  AMGN & 0.271 & 0.211 & 0.217 & 0.366 & 0.260 & 0.231 \\ 
  AMZN & 0.256 & 0.138 & 0.147 & 0.509 & 0.209 & 0.177 \\ 
  AXP & 0.062 & 0.053 & 0.052 & 0.078 & 0.061 & 0.052 \\ 
  BA & 0.061 & 0.046 & 0.050 & 0.087 & 0.058 & 0.058 \\ 
  CAT & 0.167 & 0.118 & 0.132 & 0.245 & 0.146 & 0.148 \\ 
  CRM & 0.346 & 0.156 & 0.159 & 1.043 & 0.195 & 0.176 \\ 
  CSCO & 0.121 & 0.098 & 0.094 & 0.157 & 0.106 & 0.092 \\ 
  CVX & 0.120 & 0.086 & 0.092 & 0.188 & 0.101 & 0.098 \\ 
  DIS & 0.064 & 0.042 & 0.044 & 0.111 & 0.051 & 0.048 \\ 
  GS & 0.185 & 0.132 & 0.128 & 0.289 & 0.140 & 0.127 \\ 
  HD & 0.132 & 0.099 & 0.101 & 0.198 & 0.117 & 0.108 \\ 
  HON & 0.061 & 0.053 & 0.051 & 0.075 & 0.058 & 0.051 \\ 
  IBM & 0.079 & 0.065 & 0.068 & 0.097 & 0.077 & 0.071 \\ 
  JNJ & 0.169 & 0.139 & 0.138 & 0.270 & 0.188 & 0.170 \\ 
  JPM & 0.139 & 0.090 & 0.096 & 0.245 & 0.108 & 0.105 \\ 
  KO & 0.090 & 0.065 & 0.070 & 0.136 & 0.079 & 0.077 \\ 
  MCD & 0.157 & 0.126 & 0.133 & 0.204 & 0.162 & 0.141 \\ 
  MMM & 0.062 & 0.042 & 0.044 & 0.109 & 0.052 & 0.046 \\ 
  MRK & 0.085 & 0.060 & 0.069 & 0.125 & 0.080 & 0.080 \\ 
  MSFT & 0.162 & 0.086 & 0.108 & 0.320 & 0.140 & 0.142 \\ 
  NKE & 0.117 & 0.088 & 0.091 & 0.165 & 0.104 & 0.094 \\ 
  NVDA & 0.171 & 0.116 & 0.127 & 0.270 & 0.140 & 0.139 \\ 
  PG & 0.125 & 0.082 & 0.100 & 0.205 & 0.119 & 0.123 \\ 
  SHW & 0.084 & 0.063 & 0.073 & 0.115 & 0.084 & 0.085 \\ 
  TRV & 0.087 & 0.065 & 0.070 & 0.125 & 0.080 & 0.077 \\ 
  UNH & 0.093 & 0.070 & 0.078 & 0.133 & 0.093 & 0.091 \\ 
  V & 0.172 & 0.102 & 0.104 & 0.350 & 0.136 & 0.107 \\ 
  VZ & 0.160 & 0.134 & 0.125 & 0.199 & 0.143 & 0.117 \\ 
  WMT & 0.166 & 0.135 & 0.128 & 0.235 & 0.173 & 0.126 \\ 
   \hline
\end{tabular}
\end{minipage}
\end{center}
\end{table}

\begin{table}[htbp]
\begin{center}
\begin{minipage}{250pt}
\caption{Annualized realized volatility, January 2015 - January 2026. Estimated Standard Errors of the parameter $\alpha_1$ across six estimators: Hessian (H), outer product of gradients (OPG), information matrix (IM), sandwich (S), GLM, and fully robust (FR).}\label{alpha1}
\begin{tabular}{ccccccc}
  \hline
Symbol & H & OPG & IM & S & GLM & FR \\ 
  \hline
AAPL & 0.022 & 0.017 & 0.021 & 0.031 & 0.023 & 0.027 \\ 
  AMGN & 0.022 & 0.016 & 0.021 & 0.035 & 0.025 & 0.033 \\ 
  AMZN & 0.029 & 0.020 & 0.024 & 0.046 & 0.034 & 0.034 \\ 
  AXP & 0.022 & 0.017 & 0.021 & 0.029 & 0.024 & 0.027 \\ 
  BA & 0.021 & 0.019 & 0.020 & 0.025 & 0.023 & 0.023 \\ 
  CAT & 0.021 & 0.019 & 0.020 & 0.025 & 0.023 & 0.023 \\ 
  CRM & 0.024 & 0.017 & 0.022 & 0.044 & 0.027 & 0.032 \\ 
  CSCO & 0.022 & 0.017 & 0.021 & 0.028 & 0.023 & 0.025 \\ 
  CVX & 0.021 & 0.017 & 0.020 & 0.028 & 0.023 & 0.026 \\ 
  DIS & 0.022 & 0.017 & 0.020 & 0.029 & 0.023 & 0.025 \\ 
  GS & 0.021 & 0.018 & 0.020 & 0.027 & 0.022 & 0.025 \\ 
  HD & 0.022 & 0.016 & 0.021 & 0.030 & 0.024 & 0.028 \\ 
  HON & 0.021 & 0.017 & 0.020 & 0.028 & 0.023 & 0.026 \\ 
  IBM & 0.021 & 0.016 & 0.020 & 0.029 & 0.023 & 0.027 \\ 
  JNJ & 0.022 & 0.018 & 0.021 & 0.038 & 0.029 & 0.036 \\ 
  JPM & 0.022 & 0.017 & 0.021 & 0.029 & 0.023 & 0.027 \\ 
  KO & 0.021 & 0.017 & 0.020 & 0.029 & 0.023 & 0.027 \\ 
  MCD & 0.023 & 0.018 & 0.021 & 0.030 & 0.026 & 0.026 \\ 
  MMM & 0.023 & 0.018 & 0.021 & 0.031 & 0.025 & 0.025 \\ 
  MRK & 0.021 & 0.016 & 0.020 & 0.028 & 0.023 & 0.026 \\ 
  MSFT & 0.024 & 0.017 & 0.022 & 0.036 & 0.029 & 0.029 \\ 
  NKE & 0.021 & 0.016 & 0.020 & 0.031 & 0.023 & 0.029 \\ 
  NVDA & 0.022 & 0.018 & 0.021 & 0.028 & 0.023 & 0.023 \\ 
  PG & 0.022 & 0.017 & 0.021 & 0.030 & 0.025 & 0.028 \\ 
  SHW & 0.022 & 0.018 & 0.020 & 0.027 & 0.024 & 0.023 \\ 
  TRV & 0.022 & 0.018 & 0.021 & 0.027 & 0.023 & 0.025 \\ 
  UNH & 0.022 & 0.017 & 0.021 & 0.029 & 0.025 & 0.026 \\ 
  V & 0.025 & 0.018 & 0.022 & 0.037 & 0.029 & 0.028 \\ 
  VZ & 0.021 & 0.017 & 0.021 & 0.028 & 0.024 & 0.026 \\ 
  WMT & 0.022 & 0.017 & 0.021 & 0.029 & 0.028 & 0.026 \\ 
   \hline
\end{tabular}
\end{minipage}
\end{center}
\end{table}

\begin{table}[htbp]
\begin{center}
\begin{minipage}{250pt}
\caption{Annualized realized volatility, January 2015 - January 2026. Estimated Standard Errors of the parameter $\alpha_2$ across six estimators: Hessian (H), outer product of gradients (OPG), information matrix (IM), sandwich (S), GLM, and fully robust (FR).}\label{alpha2}
\begin{tabular}{ccccccc}
  \hline
 Symbol& H & OPG & IM & S & GLM & FR \\ 
  \hline
AAPL & 0.046 & 0.026 & 0.031 & 0.083 & 0.034 & 0.037 \\ 
  AMGN & 0.043 & 0.028 & 0.035 & 0.068 & 0.042 & 0.048 \\ 
  AMZN & 0.055 & 0.031 & 0.034 & 0.103 & 0.048 & 0.040 \\ 
  AXP & 0.031 & 0.024 & 0.026 & 0.042 & 0.030 & 0.029 \\ 
  BA & 0.030 & 0.023 & 0.025 & 0.040 & 0.028 & 0.026 \\ 
  CAT & 0.036 & 0.027 & 0.029 & 0.052 & 0.032 & 0.032 \\ 
  CRM & 0.065 & 0.024 & 0.032 & 0.198 & 0.040 & 0.045 \\ 
  CSCO & 0.036 & 0.025 & 0.028 & 0.051 & 0.032 & 0.033 \\ 
  CVX & 0.045 & 0.028 & 0.032 & 0.076 & 0.035 & 0.036 \\ 
  DIS & 0.035 & 0.022 & 0.025 & 0.060 & 0.028 & 0.028 \\ 
  GS & 0.042 & 0.027 & 0.030 & 0.068 & 0.033 & 0.034 \\ 
  HD & 0.041 & 0.028 & 0.031 & 0.062 & 0.036 & 0.035 \\ 
  HON & 0.031 & 0.022 & 0.027 & 0.044 & 0.030 & 0.033 \\ 
  IBM & 0.033 & 0.025 & 0.028 & 0.044 & 0.032 & 0.034 \\ 
  JNJ & 0.040 & 0.031 & 0.032 & 0.053 & 0.044 & 0.034 \\ 
  JPM & 0.041 & 0.025 & 0.029 & 0.071 & 0.033 & 0.035 \\ 
  KO & 0.038 & 0.025 & 0.030 & 0.060 & 0.033 & 0.036 \\ 
  MCD & 0.041 & 0.031 & 0.034 & 0.056 & 0.041 & 0.038 \\ 
  MMM & 0.038 & 0.023 & 0.026 & 0.069 & 0.031 & 0.030 \\ 
  MRK & 0.030 & 0.022 & 0.025 & 0.043 & 0.029 & 0.030 \\ 
  MSFT & 0.045 & 0.027 & 0.031 & 0.082 & 0.040 & 0.036 \\ 
  NKE & 0.034 & 0.024 & 0.027 & 0.052 & 0.031 & 0.033 \\ 
  NVDA & 0.037 & 0.025 & 0.028 & 0.059 & 0.031 & 0.031 \\ 
  PG & 0.040 & 0.027 & 0.031 & 0.063 & 0.037 & 0.035 \\ 
  SHW & 0.029 & 0.023 & 0.025 & 0.037 & 0.029 & 0.027 \\ 
  TRV & 0.037 & 0.026 & 0.029 & 0.055 & 0.032 & 0.032 \\ 
  UNH & 0.031 & 0.022 & 0.026 & 0.045 & 0.032 & 0.032 \\ 
  V & 0.052 & 0.029 & 0.032 & 0.108 & 0.041 & 0.035 \\ 
  VZ & 0.040 & 0.029 & 0.032 & 0.057 & 0.036 & 0.037 \\ 
  WMT & 0.041 & 0.027 & 0.031 & 0.067 & 0.042 & 0.037 \\ 
   \hline
\end{tabular}
\end{minipage}
\end{center}
\end{table}

\begin{table}[htbp]
\begin{center}
\begin{minipage}{250pt}
\caption{Annualized realized volatility, January 2015 - January 2026. Estimated Standard Errors of the parameter $\beta_1$ across six estimators: Hessian (H), outer product of gradients (OPG), information matrix (IM), sandwich (S), GLM, and fully robust (FR).}\label{beta1}
\begin{tabular}{ccccccc}
  \hline
Symbol & H & OPG & IM & S & GLM & FR \\ 
  \hline
AAPL & 0.042 & 0.022 & 0.025 & 0.079 & 0.028 & 0.027 \\ 
  AMGN & 0.047 & 0.034 & 0.035 & 0.067 & 0.042 & 0.039 \\ 
  AMZN & 0.053 & 0.029 & 0.028 & 0.104 & 0.039 & 0.027 \\ 
  AXP & 0.021 & 0.014 & 0.016 & 0.030 & 0.018 & 0.017 \\ 
  BA & 0.019 & 0.013 & 0.013 & 0.030 & 0.015 & 0.015 \\ 
  CAT & 0.033 & 0.021 & 0.024 & 0.051 & 0.026 & 0.027 \\ 
  CRM & 0.068 & 0.023 & 0.028 & 0.216 & 0.034 & 0.035 \\ 
  CSCO & 0.031 & 0.022 & 0.022 & 0.044 & 0.025 & 0.023 \\ 
  CVX & 0.041 & 0.023 & 0.026 & 0.072 & 0.028 & 0.028 \\ 
  DIS & 0.025 & 0.013 & 0.014 & 0.047 & 0.016 & 0.015 \\ 
  GS & 0.040 & 0.024 & 0.025 & 0.069 & 0.028 & 0.028 \\ 
  HD & 0.037 & 0.024 & 0.025 & 0.062 & 0.029 & 0.029 \\ 
  HON & 0.022 & 0.016 & 0.017 & 0.033 & 0.020 & 0.020 \\ 
  IBM & 0.025 & 0.019 & 0.020 & 0.034 & 0.023 & 0.022 \\ 
  JNJ & 0.041 & 0.030 & 0.032 & 0.075 & 0.043 & 0.046 \\ 
  JPM & 0.037 & 0.020 & 0.023 & 0.070 & 0.026 & 0.027 \\ 
  KO & 0.033 & 0.020 & 0.023 & 0.056 & 0.026 & 0.028 \\ 
  MCD & 0.041 & 0.031 & 0.033 & 0.058 & 0.040 & 0.036 \\ 
  MMM & 0.027 & 0.013 & 0.015 & 0.056 & 0.018 & 0.017 \\ 
  MRK & 0.022 & 0.014 & 0.016 & 0.035 & 0.019 & 0.020 \\ 
  MSFT & 0.040 & 0.022 & 0.025 & 0.079 & 0.032 & 0.029 \\ 
  NKE & 0.028 & 0.019 & 0.020 & 0.043 & 0.022 & 0.021 \\ 
  NVDA & 0.029 & 0.017 & 0.019 & 0.050 & 0.021 & 0.021 \\ 
  PG & 0.038 & 0.022 & 0.027 & 0.068 & 0.032 & 0.035 \\ 
  SHW & 0.019 & 0.013 & 0.015 & 0.026 & 0.018 & 0.018 \\ 
  TRV & 0.030 & 0.018 & 0.021 & 0.049 & 0.024 & 0.025 \\ 
  UNH & 0.023 & 0.016 & 0.018 & 0.035 & 0.021 & 0.020 \\ 
  V & 0.050 & 0.024 & 0.026 & 0.109 & 0.034 & 0.030 \\ 
  VZ & 0.040 & 0.029 & 0.029 & 0.056 & 0.033 & 0.030 \\ 
  WMT & 0.040 & 0.027 & 0.028 & 0.066 & 0.038 & 0.033 \\ 
   \hline
\end{tabular}
\end{minipage}
\end{center}
\end{table}

\begin{table}[htbp]
\begin{center}
\begin{minipage}{250pt}
\caption{Annualized realized volatility, January 2015 - January 2026. Estimated Standard Errors of the parameter $\gamma_1$ across six estimators: Hessian (H), outer product of gradients (OPG), information matrix (IM), sandwich (S), GLM, and fully robust (FR).}\label{gamma1}
\begin{tabular}{ccccccc}
  \hline
 Symbol& H & OPG & IM & S & GLM & FR \\ 
  \hline
AAPL & 0.007 & 0.006 & 0.007 & 0.009 & 0.008 & 0.007 \\ 
  AMGN & 0.007 & 0.006 & 0.007 & 0.009 & 0.009 & 0.009 \\ 
  AMZN & 0.011 & 0.008 & 0.008 & 0.020 & 0.012 & 0.010 \\ 
  AXP & 0.005 & 0.005 & 0.005 & 0.007 & 0.006 & 0.007 \\ 
  BA & 0.005 & 0.005 & 0.005 & 0.006 & 0.006 & 0.006 \\ 
  CAT & 0.006 & 0.006 & 0.006 & 0.007 & 0.007 & 0.007 \\ 
  CRM & 0.009 & 0.007 & 0.007 & 0.017 & 0.009 & 0.008 \\ 
  CSCO & 0.006 & 0.005 & 0.006 & 0.007 & 0.007 & 0.007 \\ 
  CVX & 0.006 & 0.006 & 0.006 & 0.007 & 0.007 & 0.008 \\ 
  DIS & 0.005 & 0.004 & 0.005 & 0.005 & 0.005 & 0.005 \\ 
  GS & 0.006 & 0.006 & 0.006 & 0.006 & 0.006 & 0.006 \\ 
  HD & 0.006 & 0.006 & 0.006 & 0.007 & 0.007 & 0.007 \\ 
  HON & 0.005 & 0.005 & 0.005 & 0.006 & 0.006 & 0.006 \\ 
  IBM & 0.006 & 0.005 & 0.006 & 0.007 & 0.006 & 0.007 \\ 
  JNJ & 0.007 & 0.006 & 0.007 & 0.010 & 0.010 & 0.010 \\ 
  JPM & 0.006 & 0.006 & 0.006 & 0.008 & 0.007 & 0.007 \\ 
  KO & 0.006 & 0.006 & 0.006 & 0.006 & 0.007 & 0.006 \\ 
  MCD & 0.008 & 0.007 & 0.007 & 0.009 & 0.009 & 0.008 \\ 
  MMM & 0.005 & 0.005 & 0.005 & 0.006 & 0.006 & 0.006 \\ 
  MRK & 0.005 & 0.005 & 0.005 & 0.006 & 0.006 & 0.006 \\ 
  MSFT & 0.008 & 0.006 & 0.007 & 0.011 & 0.010 & 0.009 \\ 
  NKE & 0.006 & 0.005 & 0.005 & 0.006 & 0.006 & 0.006 \\ 
  NVDA & 0.006 & 0.006 & 0.006 & 0.007 & 0.006 & 0.006 \\ 
  PG & 0.007 & 0.006 & 0.007 & 0.008 & 0.008 & 0.008 \\ 
  SHW & 0.005 & 0.005 & 0.005 & 0.006 & 0.006 & 0.006 \\ 
  TRV & 0.006 & 0.005 & 0.006 & 0.007 & 0.007 & 0.007 \\ 
  UNH & 0.006 & 0.005 & 0.006 & 0.008 & 0.007 & 0.008 \\ 
  V & 0.009 & 0.007 & 0.008 & 0.014 & 0.010 & 0.009 \\ 
  VZ & 0.007 & 0.006 & 0.007 & 0.008 & 0.008 & 0.008 \\ 
  WMT & 0.007 & 0.006 & 0.007 & 0.009 & 0.009 & 0.009 \\ 
   \hline
\end{tabular}
\end{minipage}
\end{center}
\end{table}

\newpage
\section{Concluding remarks}\label{sec:concl}
In this work, we provide analytical derivatives of the MEM log-likelihood based on the Gamma distribution for estimation purposes, up to the second order. The implementation of an iterative algorithm to compute the MLEs of the model parameters, based on the analytical derivatives is also illustrated. Moreover, the analytical derivatives are used to provide an estimate of the asymptotic covariance matrix of the estimated parameters.  Their finite sample behaviour is compared through a simulation study under both correct specification and distributional misspecification of the data generating process. Our simulations further show that, under the considered forms of time-varying dispersion, the covariance estimator based on the GLM assumption does not generally provide a consistent estimate of the asymptotic covariance matrix, while the FR estimator remains consistent under general form of heteroskedasticity of the errors, subject to the regularity conditions required for QMLE consistency and asymptotic normality.
The iterative algorithm provided in the paper could be extended to compute the MLE of the parameters underlying the vector MEM \citep[vMEM,][]{cipollini:2017}, whereby a vector of volatility series is modelled, thus taking into account contemporaneous correlations among error terms and lagged cross-effects. \\
In the \ref{time-varying}, we provide an extension of the MEM that accommodates a time-varying dispersion parameter; a comprehensive examination of its statistical properties provides a promising direction for future research.
\newpage

The authors report generative AI was not used in their research or preparation of this manuscript‘

\appendix

\section{Univariate distributions}\label{distributions}
\subsection{Gamma Distribution}\label{gamma}
A random variable, $y$, defined on a positive support, is said to follow a Gamma distribution, namely, $y \sim \text{Ga}(\mu,\phi^2)$, if its probability density function (pdf) is given by
\begin{equation}
	f(y; \mu, \phi^2) = 
	\frac{ \left( \frac{1}{\phi^2 \mu} \right)^{\frac{1}{\phi^2}} }{ \Gamma\left( \frac{1}{\phi^2} \right) }
	y^{\frac{1}{\phi^2} - 1} \exp\left( - \frac{y}{\phi^2 \mu} \right), \quad y > 0,
	\label{eq:gamma_pdf}
\end{equation}
where $\phi^2>0$ and $\mu>0$. Notice that $\E(y)=\mu$ and $\Var(y)=\phi^2\mu^2$. The pdf of the Gamma distribution described above \citep[see also][]{rigby2019distributions} is a reparametrization of the specification provided by \cite{johnson1994continuous}, obtained by fixing $\alpha=1/\phi^2$, and $\beta=\mu\phi^2$.

\subsection{Lognormal distribution}\label{lognormal}
A random variable, $y$, defined on a positive support, is said to follow a Lognormal distribution, namely, $y \sim \text{Ln}(\mu,\phi^2)$, if its probability density function (pdf) is given by
\begin{equation}
	f(y;\mu,\phi^2) =  \frac{1}{y \sqrt{2\pi\ln(1+\phi^2) }}
	\exp\left\{- \frac{1}{2} \frac{\left[\ln\left(y\right) -\ln\left(\mu\right) -1/2\ln(1+\phi^2)\right]^2}{\ln(1+\phi^2)} \right\}, \quad y > 0,
	\label{eq:log-normal1.pdf}
\end{equation}
where $\phi^2>0$ and $\mu>0$. Notice that $\E(y)=\mu$ and $\Var(y)=\phi^2\mu^2$. The pdf of the Lognormal distribution described above is a reparametrization of the specification provided by \cite{johnson1994continuous}, obtained by fixing $\theta=0$, $\zeta=\ln(\mu) -1/2$, and $\sigma^2=\ln(1+\phi^2)$,
where $\zeta \in \mathbb{R}$ and $\sigma^2>0$ are the mean and variance of $\ln \left(y\right)$, respectively.

\section{Time-varying dispersion parameter}\label{time-varying}
Let $\left\{x_{t}\right\}$ be a non--negative time series process whose dynamics is described by the following MEM($P$,$Q$) with a time-varying dispersion parameter
\begin{align}
	&x_{t}= \mu_t\varepsilon_{t}, \ \qquad
	\varepsilon _{t}|\mathcal{I}_{t-1}\sim \mathcal{D}^+\left(1,\phi^2_t  \right), \label{errorMEMtvd} \\
	&\phi^2_t=c + \lambda \dfrac{x^2_{t-1}}{\mu^2_{t-1}},  
	\label{tvd} \\
	&\Var(x_t|\mathcal{I}_{t-1})=\phi^2_t\mu_t^2, \label{condvartvd}
\end{align} 
where $\lambda$ and $c$ are the parameters governing the dynamics of $\phi_t^2$, $\mu_t$ is specified as in \eqref{eq:mean-MEM}, and $\varepsilon_t$ is a sequence of random variables whose conditional distribution has a unit mean and time-varying dipersion parameter, $\phi_t^2$.\footnote{To ensure the nonnegativeness of $\phi_t^2$ we impose the following constraints: $c>0$ and $\lambda\geq0$.}
For simplicity, in \eqref{tvd} we assume an ARCH(1)-type dynamics, but it can be generalized to a GARCH($P$,$Q$)-type dynamics. Notice that, as stated in \cite{Barreto:2026}, the MEM in \eqref{errorMEM}-\eqref{condvar} assumes that the conditional variance is driven by the conditional mean, while here, see \eqref{condvartvd}, we consider another source of conditional volatility, related to the dynamics of the dispersion parameter. A detailed examination of the statistical properties of the model remains beyond the scope of this work and is reserved for future study.

\newpage
\newpage
\bibliography{References}

\singlespacing


\setlist[description]{font=\normalfont\itshape\textbullet\space}

\end{document}